\documentclass[12pt]{article}
\usepackage{bookmark}
\usepackage{setspace}
\usepackage{amsfonts}
\usepackage{amsmath}
\usepackage[tworuled,linesnumbered]{algorithm2e}
\usepackage{graphicx}

\usepackage[margin=2cm]{geometry}
\newcommand{\methodname}{iDTI-ESBoost} % change this name later
\newcommand{\website}
{\url{http://farshidrayhan.pythonanywhere.com/iDTI-ESBoost/}}

\newcommand{\csentence}{\textbf{1}}

\title{{\methodname}: Identification of Drug Target Interaction Using Evolutionary and Structural Features with Boosting}
\author{Farshid Rayhan$^1$, Sajid Ahmed$^1$,  Swakkhar Shatabda$^1$\thanks{Corresponding email: swakkhar@cse.uiu.ac.bd}, Dewan Md Farid$^1$,\\ Zaynab Mousavian$^3$, Abdollah Dehzangi$^4$ and M Sohel Rahman$^2$, 
	\\
	$^1$Department of Computer Science and Engineering, United International University\\
	$^2$Department of Computer Science and Engineering, \\Bangladesh University of Engineering and Technology\\
	$^3$ Laboratory of Systems Biology and Bioinformatics (LBB), \\Institute of Biochemistry and Biophysics, University of Tehran, Tehran, Iran\\
	$^4$ Carver College of Medicine, University of Iowa
}
\date{}

\begin{document}
\maketitle
\doublespacing
\begin{abstract}
\noindent\textbf{Background:}\\
Prediction of new drug-target interactions is extremely important as it can lead the researchers to find new uses for old drugs and to realize the therapeutic profiles or side effects thereof. However, experimental prediction of drug-target interactions is expensive and time-consuming.  As a result, computational methods for prediction of new drug-target interactions have gained much interest in recent times.\\
\noindent\textbf{Results:}\\
We present {\methodname}, a prediction model for identification of drug-target interactions using evolutionary and structural features. Our proposed method uses a novel balancing technique and a boosting technique for the binary classification problem of drug-target interaction. On four benchmark datasets taken from a gold standard data, {\methodname} outperforms the state-of-the-art methods in terms of area under Receiver operating characteristic (auROC) curve. {\methodname} also outperforms the latest and the best-performing method in the literature to-date in terms of area under precision recall (auPR) curve. This is significant as auPR curves are argued to be more appropriate as a metric for comparison for imbalanced datasets, like the one studied in this research. In the sequel, our experiments establish the effectiveness of the classifier, balancing methods and the novel features incorporated in {\methodname}.\\
\noindent\textbf{Conclusions:}\\
{\methodname} is a novel prediction method that has for the first time exploited the structural features along with the evolutionary features to predict drug-protein interactions. We believe the excellent performance of {\methodname} both in terms of auROC and auPR would motivate the researchers and practitioners to use it to predict drug-target interactions. To facilitate that, {\methodname} is readily available for use at: {\website}
\end{abstract}

\section*{Introduction}
Target specific drug design is one of the key techniques in therapeutic drug discovery \cite{keiser2009predicting}. Prediction of new drug target interactions can lead the researchers to find new uses for old drugs and to realize the therapeutic profiles or side effects thereof \cite{cheng2013prediction,wu2017sdtnbi,campillos2008drug}. Since experimental prediction of drug-target interaction is expensive and time-consuming \cite{haggarty2003multidimensional,kuruvilla2002dissecting}, computational methods have been gaining increasing popularity in recent years.

The computational approaches taken in the literature to address the drug-target interaction include, but are not limited to ligand based methods \cite{hopkins2014role,keiser2007relating}, target or receptor based methods \cite{ma2013drug,pan2013molecular}, gene ontology based methods \cite{mutowo2016drug}, literature text mining methods \cite{plake2011computational,zhu2005probabilistic}, etc. The performance of the ligand based methods degrade as
the number of known ligands of a particular target protein decreases. Receptor based methods often use docking simulation \cite{morris2009autodock4} and hugely rely on availability of the three dimensional structure of the protein targets. Notably, finding three-dimensional structures of the proteins is a expensive and time-consuming task using NMR and X-ray Crystallography. Moreover, three dimensional structures are very difficult to predict for ion channel proteins and G-protein coupled receptors
(GPCRs). On the other hand, the tremendous growth in the Biomedical literature has increased the problem of redundancy in the compound/gene names as the main obstacle for literature based systematic text mining methods.

Recently, chemo-genomic methods \cite{mousavian2014drug} have been attempted to predict drug-target interactions. These methods are mainly learning-based methods \cite{yamanishi2008prediction,bleakley2009supervised}, graph-based methods \cite{wang2013drug,chen2012drug} and network-based methods \cite{alaimo2013drug,cheng2012prediction}. In the supervised learning setting, several classification algorithms have been used in the literature. Examples include support vector machine \cite{mousavian2016drug,keum2017self}, deep learning \cite{chan2016large},	fuzzy \cite{xiao2013icdi}, nearest neighbor algorithm \cite{he2010predicting}, etc. Yamanishi et al. \cite{yamanishi2008prediction} proposed a formalization of the drug–target interaction inference as a supervised learning problem for a bipartite graph. In that pioneering work, they also proposed a gold standard dataset that had been later used extensively in the literature \cite{mousavian2016drug,yamanishi2010drug,chan2016large}. In a subsequent work, the same authors \cite{yamanishi2010drug} explored the relationship among the chemical space, the pharmacological space and the topology of drug-target interactions networks and applied distanced based learning.  Wang et al. proposed RLS-KF \cite{hao2016improved} that uses regularized least squares method integrated with nonlinear kernel fusion.
Drug-based similarity inference (DBSI) was proposed in \cite{cheng2012prediction} utilizing two dimensional chemical structural similarity. Another method, KBMF2K, was proposed in \cite{gonen2012predicting} that used chemical and genomic
kernels and bayesian matrix factorization. Among other methods, NetCBP \cite{chen2013semi}, DASPfind \cite{ba2016daspfind}, SELF-BLM \cite{keum2017self}, etc. are noteworthy. In a recent work \cite{mousavian2016drug}, position specific scoring matrix based bigram features and molecular fingerprint were used to predict drug target interactions. In the absence of three dimensional structures of the protein target, most of the supervised learning methods in the literature do not exploit the structure based features.

In this paper, we present {\methodname}, a method for {\bf i}dentification of {\bf D}rug {\bf T}arget {\bf I}nteraction Using {\bf E}volutionary and {\bf S}tructural Information with {\bf Boost}ing. We exploit the structural features along with the evolutionary features to predict drug-protein interactions. Our work was inspired due to the modern successful secondary structural prediction tools like SPIDER2 \cite{yang2016sixty,yang2017spider2} and its use  to generate features in supervised learning and classification \cite{lopez2017sucstruct}. Our proposed method uses a novel set of features extracted using structural information along with the evolutionary features and molecular fingerprints of drugs. To handle the large amount of imbalance in the data, we propose a novel balancing method and use it along with a boosting algorithm to achieve superior performance over the state-of-the-art algorithms. In our experiments our method, {\methodname}, has shown to have significantly outperformed other methods on a gold standard data set widely used in the literature under standard evaluation criteria. Our method is publicly available to use at: {\website}. 

The rest of the paper follows are the general suggestions made in \cite{chou2011some}: description of dataset, formulation of statistical samples, selection and development of a powerful classification algorithm, demonstration of the performance of the predictor using cross-validation, implementation of web server followed by a conclusion. 

\section*{Materials and Methods}

\begin{figure}[h!]
\begin{center}
\includegraphics[width=0.45\textwidth]{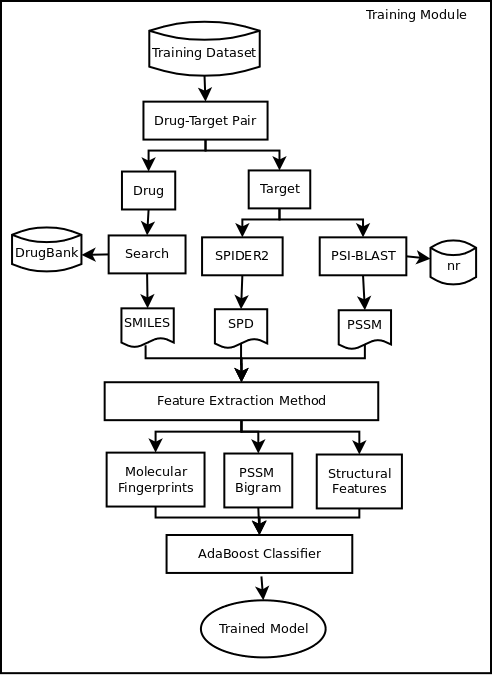}
\end{center}
\caption{\csentence{Schematic Diagram.} Schematic diagram of the training module of {\methodname} showing the steps of the training phase. \label{figTraining}}
\end{figure}

In this section, we provide the details of the benchmark datasets, feature extraction and balancing methods, classifiers and evaluation metrics used in this research work. Figure~\ref{figTraining} depicts the training module of our proposed method, {\methodname}. The training dataset of {\methodname} contains both interacting (positive) and non-interacting drug-target pairs. For each instance of drug-target pair, a drug is searched in the DrugBank database \cite{knox2011drugbank} to fetch the drug chemical structure in SMILES format. Similarly, a target protein sequence is first fetched from KEGG database %}{This step is NOT presented in the flowchart!}
 \cite{kanehisa2000kegg} and then fed to SPIDER2 \cite{yang2017spider2} and PSI-BLAST \cite{altschul1997gapped} in order to receive, respectively, structural information as an SPD file and position specific scoring matrix (PSSM) based profile containing evolutionary information. A feature extraction module then uses these files to generate three types of features: drug molecular fingerprints, PSSM bigram and structural features based on the output of the secondary structure prediction software SPIDER2. Features generated in this phase is then fed to an AdaBoost classifier that learns the model for prediction purposes.

The prediction module is very similar to that of the training module shown in Figure~\ref{figTraining}. For prediction, a query drug-target pair is feed to the system in a similar way to extract three types of features and then the trained and stored model is used to predict whether the given drug-target pair is interacting or non-interacting.

\subsection*{Drug-target Interaction Datasets}
In this paper, we have used the gold standard datasets introduced by Yamanishi et al. in \cite{yamanishi2008prediction}. The datasets are publicly available at: \url{http://web.kuicr.kyoto-u.ac.jp/supp/yoshi/drugtarget/}. Yamanishi et al. used DrugBank \cite{wishart2008drugbank}, KEGG BRITE \cite{kanehisa2008kegg}, BRENDA \cite{schomburg2004brenda} and SuperTarget \cite{gunther2008supertarget} to extract information about drug-target interactions. They used the known drugs to four types of protein targets, namely, enzymes, ion channels, g-protein coupled receptors (GPCRs) and nuclear receptors. The number of proteins in these classes are 664, 204, 95 and 26 respectively, that interact with, respectively, 445, 210, 223 and 54 drugs through 2926, 1476,635 and 90 known interactions. A brief description of the datasets are given in Table~\ref{tabDataset}. These benchmark datasets have been used in many studies in the literature \cite{mousavian2016drug,yamanishi2010drug,cheng2012prediction,chan2016large} and are referred to as the `gold' standard.

\subsection*{Graph Construction from the Dataset}
%\note{Note that I have changed the title. Is this okay? I have also made some notational changes to make things simpler.}
Based on the interactions of four types of proteins with known drugs, we build positive and negative samples for each dataset using a method similar to the one used in \cite{mousavian2016drug} as follows. The drug-target interaction network for each dataset is a bipartite graph, $G=(V,E)$, where the set of vertices is $V=D\cup T$ such that $D$ is the set of drugs and $T$ is the set of targets, $D\cap T = \emptyset$ and the set of edges is $E$. Here, any edge $e = (d,t)\in E$ denotes an interaction only between a drug, $d \in D$ with a protein target, $t \in T$. Now, for a particular graph from a dataset, all the known interactions in the graph represented by its edges are considered to be positive samples and the non-existent edges are taken as negative samples. Note that, here, non-existent edges refer to the possible valid edges only that are not there; i.e., they do not include edges among the vertices of the same partite set. Formally, a dataset is an union of positive and negative sets as follows:

\begin{equation}
\mathbb{S}=\mathbb{S}^+ \cup \mathbb{S}^-
\end{equation}

Here, $\mathbb{S}^+=\{(u,v): u \in D, v \in T, (u,v) \in E \}$, and $\mathbb{S}^-=\{(u,v): u \in D, v \in T, (u,v) \notin E \}$. For example, in the nuclear receptor, there are 54 drugs and 26 proteins with possible $54\times 26 = 1404$ interactions. Since 90 interactions are known, these are treated to be positive and the rest 1314 as negative. The same procedure was followed for each of the datasets. As expected, the constructed datasets using this technique are imbalanced as the number of negative samples far outnumbers that of positive samples. This issue is attended to later by applying some balancing techniques.
\subsection*{Feature Extraction}
A dataset constructed in this way has drug-target pairs as instances. In the feature extraction phase, a drug identifier is looked up in the KEGG databased \cite{kanehisa2008kegg} and the corresponding SMILES format is downloaded from the DrugBank database \cite{knox2011drugbank}. The features based on drugs are generated using this SMILES data.

Similarly, a protein target of each pair is first searched with in the KEGG database \cite{kanehisa2008kegg} to fetch the protein sequence. This protein sequence is then fed to two different software: Position Specific Iterated BLAST (PSI-BLAST) \cite{altschul1997gapped} to fetch evolutionary profile based position specific scoring matrix (PSSM) and a secondary structure prediction tool, SPIDER3 \cite{yang2017spider2} to generate SPD3 files that contains the structural information. Three groups of features are extracted using these three files. The details are described in the rest of this section.

\subsubsection*{SMILES Based Features}
Several descriptors are used to represent the features or properties of drug compounds \cite{todeschini2008handbook}. To this end, one of the most popular features is molecular fingerprints, which is widely used for similarity searching \cite{tabei2013scalable}, clustering \cite{tabei2012identification}, and classification \cite{mousavian2016drug}. Each drug compound is represented by 881 chemical substructures defined in PubChem database \cite{chen2009pubchem}. The presence (absence) of a particular substructure is encoded as 1 (0). Thus the length of this molecular fingerprint based feature is 881. We used the \textit{rcdk} package of R \cite{guha2007chemical} to extract these molecular fingerprints based features.

\subsubsection*{PSSM Based Features}
We used the PSSM matrix returned by the PSI-BLAST software to generate evolutionary features from the protein target sequences. Each PSSM file contains a PSSM matrix that is constructed after multiple sequence alignment using the non redundant (NR) database. The PSSM file contains a matrix $M$ of dimension $L \times 20$, where $L$ is the length of the protein and each of the entries in this matrix, $m_{ij}$, represents the probability of observing the $j$-th amino acid in the $i$-th location of the given protein sequence. We first convert this matrix $M$ to a normalized matrix using a normalization technique similar to that proposed in \cite{sharma2015predict}. The dimension of this matrix is same as the original matrix $M$. After that we generate PSSM-bigram features using the following equation:

\begin{equation}
\text{PSSM-bigram} (k,l) = \frac{1}{L}\sum_{i=1}^{L-1}N_{i,k}N_{i+1,l}~ (1\le k \le 20, 1\le l \le 20)
\end{equation}

Bigram features for PSSM were first proposed in \cite{paliwal2014tri}  and subsequently used successfully in drug-target interaction prediction in \cite{mousavian2016drug}. Total number of features generated usign this method is 400.
\subsubsection{Structure Based Features}
The traditional drug discovery is a lock-key problem, where the lock is the target. The structure of the target thus play a very important role in traditional drug discovery and is at the center of the docking based software. We make a hypothesis that even if the full structure is not present for the targets, estimated structural properties thereof can play an important role in drug-target interaction prediction. Structural features are generated using the structural information generated and stored in SPD files by SPIDER2 software. The information generated by SPIDER2 are: accessible surface area (ASA), secondary structural (SS) motifs, torsional angles (TA) and structural probabilities (SP). Following features are generated using these information:

\begin{enumerate}
\item {\bf Secondary Structure Composition:} This feature is the normalized count or frequencies of the structural motifs present at the amino-acid residue positions. There are three types of motifs: $\alpha$-helix (H), $\beta$-sheet (E) and random coil (C). SPIDER2 returns a vector $SS$ of dimension $L\times 1$ containing this information. Thus we can define this feature as following:
\begin{equation}
\text{SS-Composition}(i) = \frac{1}{L}\sum _{j=1}^{L} c_{ij},  1 \le i \le 3
\end{equation}

Here, $L$ is the length of the protein and $$c_{ij}=\begin{cases}
1, \text{ if } SS_j = f_i\\
0, \text{else}
\end{cases}$$

Here, $SS_j$ is the structural motif at position $j$ of the protein sequence and $f_i$ is one of the 3 different motif symbols.

\item {\bf Accessible Surface Area Composition:} The accessible surface area composition is the normalized sum of accessible surface area defined by:

\begin{equation}
\text{ASA-Composition}=\frac{1}{L}\sum_{i=1}^{L}ASA(i)
\end{equation}
Here ASA is the vector of accessible surface area of dimension $L\times 1$ containing the values of accessible surface area for all the amino acid residues.
\item {\bf Torsional Angles Composition:}
Four different types of torsional angles: $\phi$, $\psi$, $\tau$ and $\theta$ are returned by SPIDER2 for each residue. First, we convert each of them into radians from degree angles and then take sign and cosine of the angles at each residue position. Thus we get a matrix of dimension $L\times 8$. We denote this matrix by $T$. Torsional angles composition is defined as:
\begin{equation}
\text{TA-Composition(k)}= \frac{1}{L}\sum_{i=1}^{L}T_{i,k} ~(1\le k \le 8)
\end{equation}

%\item {\bf Structural Probabilities Composition:} Structural probabilities for each position of the amino-acid residue are given in SPD3 file as a matrix of dimension $L \times 3$. We denote it by $P$. Structural probabilities composition is defined as:
%\begin{equation}\text{SP-Composition(k)}= \frac{1}{L}\sum_{i=1}^{L}P_{i,k} (1\le k \le 3)\end{equation}

\item {\bf Torsional Angles Bigram:}
The Bigram for the torsional angles is similar to that of the PSSM matrix and is defined as:

\begin{equation}
\text{TA-bigram} (k,l) = \frac{1}{L}\sum_{i=1}^{L-1}T_{i,k}T_{i+1,l}~ (1\le k \le 8, 1\le l \le 8)
\end{equation}

\item {\bf Structural Probabilities Bigram:} Structural probabilities for each position of the amino-acid residue are given in the SPD2 file as a matrix of dimension $L \times 3$, which we denote by $P$. Recall that, there are three types of structiral motifs, namely, $\alpha$-helix (H), $\beta$-sheet (E) and random coil (C). The Bigram of the structural probabilities is similar to that of PSSM matrix and is defined as:

\begin{equation}
\text{SP-bigram} (k,l) = \frac{1}{L}\sum_{i=1}^{L-1}P_{i,k}P_{i+1,l}~ (1\le k \le 3, 1\le l \le 3)
\end{equation}
\item {\bf Torsional Angles Auto-Covariance:}
This feature is also derived from the torsional angles and is defined as: %\note{A few lines to explain this feature would be fruitful. Also, what is DF in the formula in this one and the next one?}

\begin{equation}
\text{TA-Auto-Covariance} (k,j) = \frac{1}{L}\sum_{i=1}^{L-k}T_{i,j}T_{i+k,j} ~ (1\le j \le 8, 1\le k \le DF)
\end{equation}
This feature group depends on parameter DF which is the distance factor. In this study, we used DF = 10.
\item {\bf Structural Probablities Auto-Covariance:}
This feature is also derived from the structural probabilities and is defined as:

\begin{equation}
\text{SP-Auto-Covariance} (k,j) = \frac{1}{L}\sum_{i=1}^{L-k}P_{i,j}P_{i+k,j} ~ (1\le j \le 3, 1\le k \le DF)
\end{equation}
\end{enumerate}

 A brief summary of the three group of features derived from each drug-target pair is given in Table~\ref{tabFeatSum}. %\note{One question that came to my mind is why these features have been described in this particular order. For example, intuitively, I would probably go for Tprsional Angles related features in one cluster. Later I came to realize that this relates to the combinations used in the experiments. But even then here the order should perhaps be ``natural". Think about it.}

\subsection*{Balancing Methods}
Recall that, each of our four datasets is heavily imbalanced. Several sampling techniques in the literature have been deployed in imbalanced settings of data: random under sampling \cite{mousavian2016drug}, synthetic over sampling \cite{chawla2002smote} , balanced random sampling (BRS) \cite{yu2010simple}, neighborhood cleaning rule \cite{laurikkala2001improving}, cluster based under sampling \cite{yen2009cluster, rahman2013cluster}, etc. In this paper, we explore random under sampling (RUS) method as done previously for drug-target interaction prediction in \cite{mousavian2016drug}. We also propose a novel modified cluster based under sampling method based on \cite{rahman2013cluster} as follows. In this method, the dataset is first divided into two subsets as major class and minor class . In the major class $k$-means clustering is applied to divide the major class samples in $k$ clusters . But the minor class samples are kept unchanged. Now from the $k$ clusters of major class samples, subsamples are chosen randomly to represent the entire major class. We denote this method as \textit{cluster based under sampling (CUS)} throughout this paper. The random under sampling will be denoted as \textit{random under sampling (RUS)}. The pseudo-code for the CUS algorithm is given in Algorithm~\ref{algCUS}.

\begin{algorithm}[!htb]
\DontPrintSemicolon
$major,minor\gets$divide$(dataset)$\;
$clusters \gets k$-MeansClustering$(major,k)$\;
$kCombined \gets \emptyset$\;
\For{ each cluster $\in clusters$ }{
 $clusteredData \gets major.getData(cluster)$\;
 $reduced \gets randomSubsample(clusteredData,h)$\;
 $kCombined \gets kCombined \cup reduced$\;
}
$dataset'\gets kCombined \cup minor$\;
\Return $dataset'$
\caption{ClusterBasedUnderSampling($dataset$,$k$,$h$)}
\label{algCUS}
\end{algorithm}

Our CUS algorithm depends on two parameters, namely, $k$ and $h$. In our experiments, we have varied $k$ for values from $5\cdots 30$ and found the the best performing value to be 23. However, more sophisticated clustering algorithms can be applied on this data. The role of the parameter $h$ is to control the random under sampling of the clustered majority class samples.

\subsection*{Description of the classifier}
We have selected the adaptive boosting algorithm (AdaBoost) \cite{freund1995desicion} as our classification algorithm. Adaptive boosting is a meta or ensemble classifier that uses several weak learning algorithms or weak classifiers and improves over their performance. We choose decision tree classifiers as the weak classifiers. AdaBoost is a meta-classifier of the following form:
\begin{equation}
g = \sum_{t=1}^{T}\alpha_t h_t(x)
\end{equation}

AdaBoost iteratively adds up a weak classifier $h_t(x)$ at each iteration of the algorithm weighted by $\alpha_t$ where $\alpha_t$ is the weight achieved from the error function $\epsilon_t$ for the weak classifier $h_t(x)$ at iteration $t$. Each of these weak classifiers is chosen in a way so as to minimize the error on the training sample weighted by the distribution $D_t$:
\begin{equation}
h_t \in \underset{h\in H}{argmin} \underset{i\sim D_t}{\text{Pr}}[h_t(x)\neq y_i] = \underset{h \in H}{argmin} \sum_{i=1}^{m}D_t1_{h_t(x)\neq y_i}
\end{equation}

The algorithm of AdaBoost \cite{freund1995desicion} is sketched in Algorithm~\ref{algADA} following the notations of \cite{mohri2012foundations}.

\begin{algorithm}[!htb]
\DontPrintSemicolon
\For{ $i\gets 1$ to $m$ }{
 	$D_1\gets \frac{1}{m}$\;
}
\For{ $i\gets 1$ to $T$ }{
 $h_t \gets $ decision tree classifier with small error $\epsilon_t={\text{Pr}}[h_t(x)\neq y_i]$\;
 $\alpha_t	\gets \frac{1}{2}\text{log}\frac{1-\epsilon_t}{\epsilon_t}$\;
 $Z_t\gets 2[\epsilon_t(1-\epsilon_t)]^{\frac{1}{2}}$\;
 \For{$i\gets 1$ to $m$}{
 $D_{t+1}(i) \gets \frac{1}{Z_t}D_t e^ {-\alpha_ty_ih_t(x_i)}$
 }
}
$g \gets \sum_{t=1}^{T}\alpha_th_t$\;
\Return $h=sign(g)$
\caption{AdaBoost($dataset = (X,Y)$)}
\label{algADA}
\end{algorithm}

\subsection*{Performance Evaluation}
A large variety of performance metrics are used in the literature to compare the performance of supervised learning methods \cite{powers2011evaluation}. The gold datasets that is used in the literature of drug-target interaction prediction is largely imbalanced and the number of negative samples largely outnumbers that of the positive samples. Therefore, the typical measures like accuracy does not make much sense. Moreover, the output of the classifier generating probabilistic outputs depends on the thresholds or the values predicted by it for each of the predicting classes. In such cases, thresholds or values play and important role on the sensitivity and specificity of the classifiers. Two measures that are independent of the values or thresholds set for decision making are area under curve for Receiver Operating Characteristic (auROC) and area under precision recall curve (auPR). These two measures are widely used in the literature of drug-target interaction prediction \cite{mousavian2016drug,chen2013semi,cao2012large,chan2016large} and thus have become standard metrics for comparison.

Lets assume, $P$ is the total number of positive samples in a dataset and $N$ is the total number of negative samples in a dataset. Let $TN,TP,FN,FP$ denote the number of true positives, true negatives, false negatives and false positives predicted by a classifier. True positives (negatives) are correctly classified positive (negative) samples by the classifier. Conversely, false positives (negatives) are negative (positive) samples incorrectly predicted as positives (negatives) by the classifier. Following these notions, we can define \emph{sensitivity} or \emph{true positive rate} as follows:
\begin{equation}
Sensitivity = \frac{TP}{TP+FN}
\end{equation}
Therefore, sensitivity is the ratio of correctly predicted positive samples to the total number of positive samples. \emph{Precision} is defined as the positive predictive rate (PPV) as follows:

\begin{equation}
Precision  = \frac{TP}{TP+FP}
\end{equation}

Therefore, precision shows the percentage of positive predictions by the classifiers that are accurate. Another important measure is specificity (SPC) or true negative rate defined as follows:
\begin{equation}
SPC = \frac{TN}{TN+FP}
\end{equation}

Fall-out or false positive rate (FPR) is the ration of the number of wrongly classified negative samples to the total number of negative samples defined as follows:
\begin{equation}
FPR = \frac{FP}{FP+TN} = 1-SPC
\end{equation}

F1 Score is the harmonic mean of the precision and sensitivity and defined as follows:

\begin{equation}
F1=\frac{2TP}{2TP+FP+FN}
\end{equation}

All theses  performance measures have values with in the range $[0 \cdots 1]$, 0 being the worst and 1 being the best. 

%\note{Is everything defined above used in our analysis?}

Another score that is often used in comparison is called %\annote{Mathew's Correlation Coefficient (MCC)}{We have not reported it in our analysis, have we?} 
defined as follows:
\begin{equation}
MCC = \frac{(TP\times TN)-(FP\times FN)}{\sqrt{(TP+FP)(TP+FN)(TN+FP)(TN+FN)}}
\end{equation}
Value of this coefficient ranges from $-1$ to $+1$, where $+1$ means a perfect predictor and $-1$ means a total disagreement.

Receiver operating characteristic (ROC) curve plots true positive rate against false positive rate at various threshold values. The performance of a predictor is calculated by the area under the ROC curve (auROC). A perfect classifier have a auROC value of 1 and a random classifier have a value of 0.5. However, for imbalanced datasets like ours, area under precision recall curve (auPR) is of more significance \cite{mousavian2016drug} as follows: auPR curve plots the precision rate vs the recall rate at different threshold values. This score penalizes the false positives more as compared to auROC and thus more suitable for skewed datasets. The value of auPR ranges from $0$ to $1$ and the higher the value is the better. 
%\note{For auPR curve, please mention the range of values identifying which values indicate good vs bad}

It is very important to test the methods to check and balance the bias-variance trade-off \cite{friedman1997bias}. Various methods of sampling are used to measure the performance of supervised learning algorithms \cite{efron1983leisurely}. Among them mostly used are $k-$fold cross validation and jack knife tests. Because of the high imbalance, dimensionality and cardinality of the datasets, in most of the methods in the literature, 5-fold cross validation have been preferred and used as the sampling method  \cite{cao2012large,chan2016large,mousavian2016drug,chen2013semi}. We also use the 5-fold cross validation to test our method for the sake of fair comparison with the other state-of-the-art methods.

In the 5-fold cross validations, first the dataset is randomly split into five equal parts retaining the ratio of imbalance in each split same to the original dataset. Each time one part of the dataset is used as test and the other four are used as training data. First the balancing techniques are applied to the training data (clustered or random) and then the classifier is used to train the data into a model. The model is then used to predict the labels for the test data. Thus all the drug-target pairs in the datasets are used in testing the classifier performance using cross-validation. The measures reported are the average of all 5-fold results.

\section*{Results and Discussion}
In this section, we present the results of {our} experiments. All the methods were implemented in Python language using Python3.4 version and Scikit-learn library \cite{pedregosa2011scikit} of Python was used for the implementation of the machine learning algorithms. All experiments were conducted on a Computing Machine hosted by CITS, United International University. %The machine was equipped with 8 core processors each core having a Dell R 730 Intel Xeon Processor (E5-2630 V3) with 2.4 GHz speed and 18.5 GB of memory. 
Each of the experiments was carried out 5 times %}{Seems to low! Or is it the norm for drug target interaction studies?} 
and the average is reported as the results. We perform several types of experiments. In particular, we conduct four different sets of experiments as follows. First we investigate the effectiveness of the different feature groups as mentioned in Table \ref{tabFeatSum}. Recall that, in Table \ref{tabFeatSum}, four different feature groups, namely, A, B, C and D, were formed. Secondly, we conduct experiments to investigate the effectiveness of the classifiers used in our research. Subsequently, we also experiment the effectiveness of the balancing methods applied on our highly imbalanced datasets. Finally, we also conduct experiments to test our method, {\methodname}, against the state-of-the-art.

\subsection*{Effectiveness of Feature Groups}
We created four different feature groups to see the effects of the different sets of features on the classifier performance. The feature groups have already been reported in Table~\ref{tabFeatSum}. Group A contains 1281 features and was previously used in \cite{mousavian2016drug}. We further added other groups, namely, B, C and D, incrementally in that order with the base feature group i.e., Group A and achieve features of size 1293, 1403 and 1476 respectively. We have performed two sets of experiments to test the effectiveness of the feature groups. In both of experiments we varied the feature groups and ran different classifiers and applied different balancing methods on the data to analyze the effect. Results of these experiments are reported in Table~\ref{tabClassifiers} and Table~\ref{tabFeatBalance}.

Table~\ref{tabClassifiers} reports the performance of three different classifiers on the four datasets during our experiments. Note that, though this experiment was intended for classifier selection, we clearly see that the best results in terms of auPR and auROC were found only when the structural features are added. For enzymes dataset, the best result in terms of auPR was 0.66 found with the combination A,B,C which is using structural composition and structural auto-covariance groups with PSSM-bigram and molecular fingerprint based features. It was slightly better then the case when we use all the features A,B,C,D  and got auPR of 0.66. In terms of auROC, the results were somewhat comparable to each other; however, the best result was achieved when all the four feature groups were used in combination. Thus enzyme dataset shows the effectiveness of structural information based features.

Datasets ion channels and GPCRs showed similar performance in terms of auPR. Nuclear receptors showed highest auPR value when only the composition features, i.e., Group B were added with the base features. %\note{One issue that a reviewer might ask is as follows: why did you only consider, A vs A+B vs A+B+C vs A+B+C+D; why did not you consider different combinations? This question especially pops up here, as you are saying A+B is good but not A+B+ ....; so, how about A+C or may be A+B+D, leaving out only C?} 
The increase in the value of auROC clearly reveals the effectiveness of the structural features (Groups B,C,D) when added to the base feature (Group A).

The next set of experiments were run to show the performance of different balancing or under sampling methods in the training data using various feature groups. These results are shown in Table~\ref{tabFeatBalance}. These experiments were run using the AdaBoost classifier. The results in Table~\ref{tabFeatBalance} clearly shows that for all the datasets, the best results in terms of auPR and auROC were found when structural features have been added.  In case of the GPCRs, the auPR was found to be the highest at 0.5 when three feature groups, namely, Groups A,B, and C have been combined. Apart from this, in all other datasets, the all four groups combined have shown superior performance both in terms of auPR and auROC. Our hypothesis that the added structural features play an important role in the prediction of drug-target interaction is thus justified  according to these experiments.

\subsection*{Effectiveness of the AdaBoost Classifier}
\begin{figure}[h!]
\begin{center}
\begin{tabular}{cc}
\includegraphics[width=0.45\textwidth]{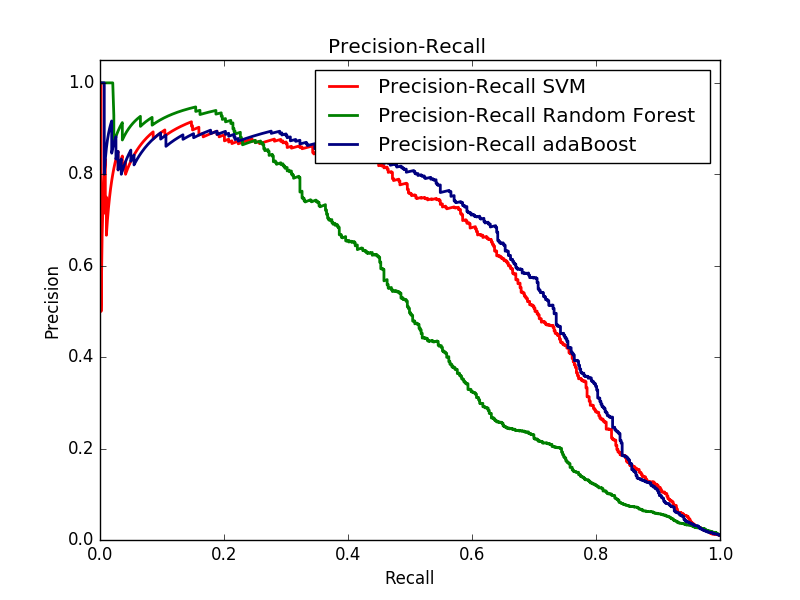}&
\includegraphics[width=0.45\textwidth]{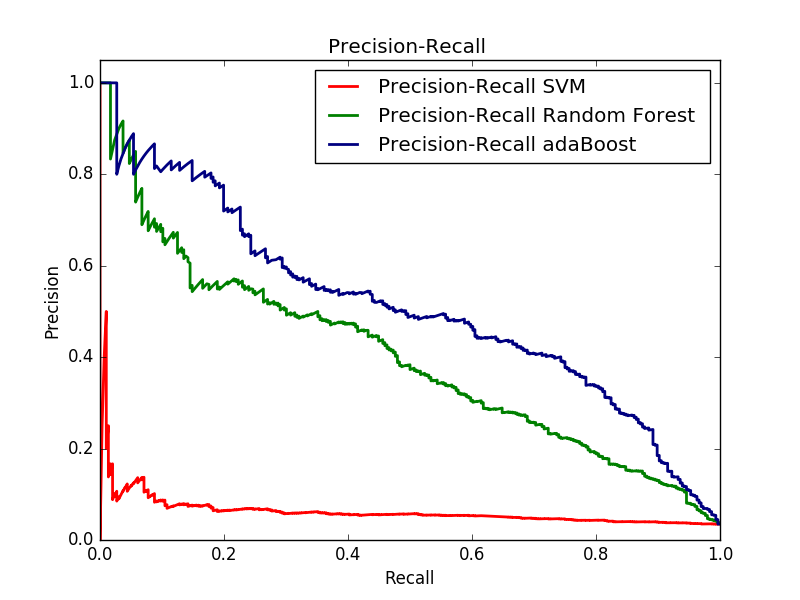} \\
(a)&(b)\\
\includegraphics[width=0.45\textwidth]{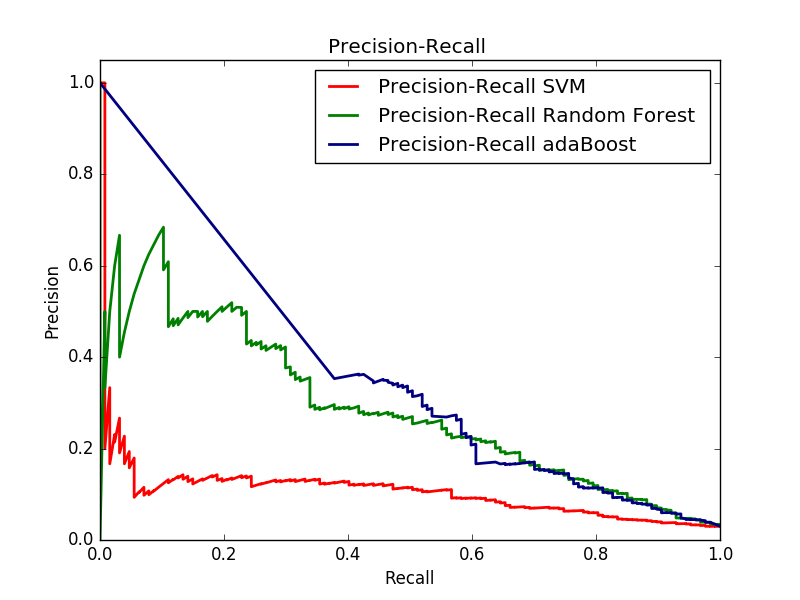}&\includegraphics[width=0.45\textwidth]{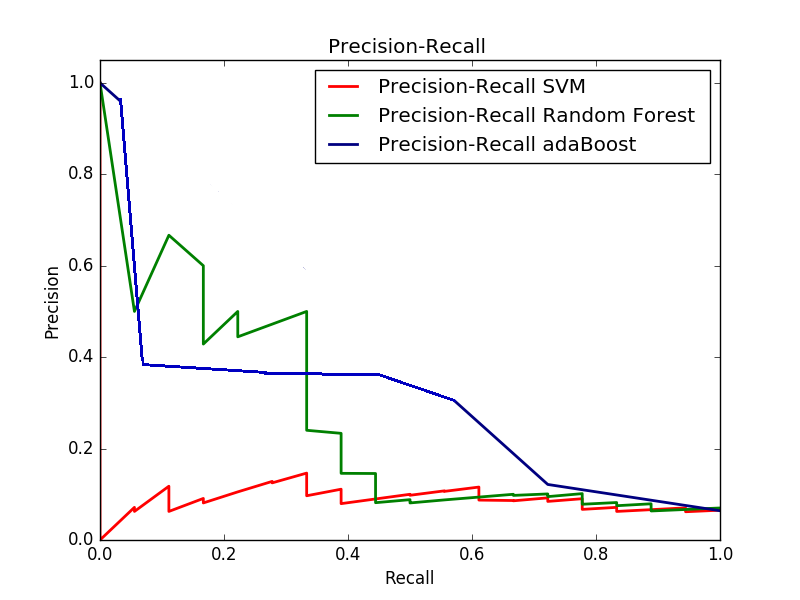} \\
(c)&(d)\\
\end{tabular}
\end{center}
\caption{\csentence{Classifier auPR Analysis.} Precision-Recall curves of different classifier algorithms using random under sampling and all the feature combinations on four datasets: (a) enzymes (b) ion channels (c) GPCRs (d) nuclear receptors.\label{figCLSPR}}
\end{figure}

To test and select the suitable classifier for our problem, we test three different classifiers: AdaBoost ensemble classifier \cite{freund1995desicion} with decision tree as the weak classifier, Random Forest \cite{ho1998random} and Support Vector Machines \cite{cortes1995support}. For these experiments, we used random under sampling as the balancing method. As features, four different combinations were used as has been mentioned already. The results in terms of auPR and auROC are presented in Table~\ref{tabClassifiers}. Here for each of the datasets and feature groups combinations bold faced values in the table represents the highest values achieved for that combination. It is evident that except for one case in the enzymes dataset, AdaBoost classifier has shown superior performance in terms of auPR across all feature groups combinations. It is also worth-noting that for all datasets, the highest auPR value was achieved by AdaBoost. The precision-recall curves for these experiments across all feature groups combinations are illustrated in Figure~\ref{figCLSPR}.

%\note{Why Figure 3 is discussed before Figure 2?}

\begin{figure}[h!]
\begin{center}
\begin{tabular}{cc}
\includegraphics[width=0.45\textwidth]{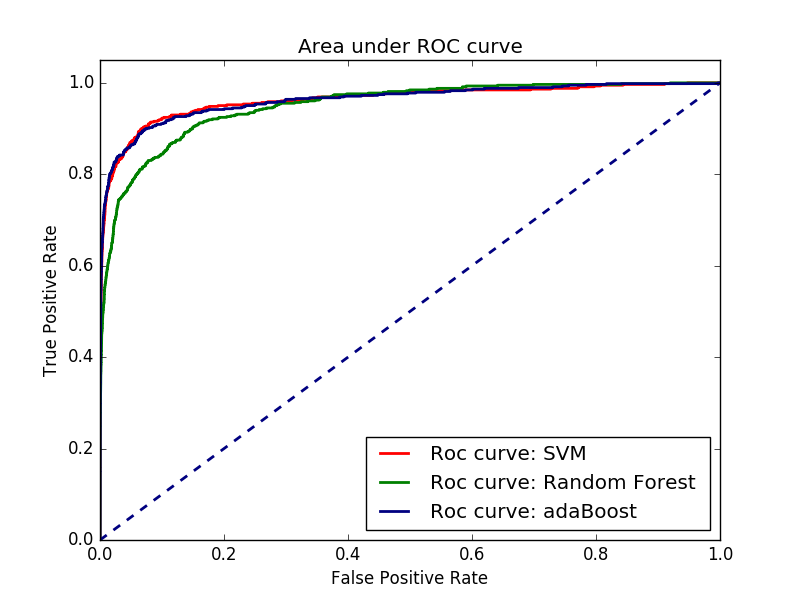}&
\includegraphics[width=0.45\textwidth]{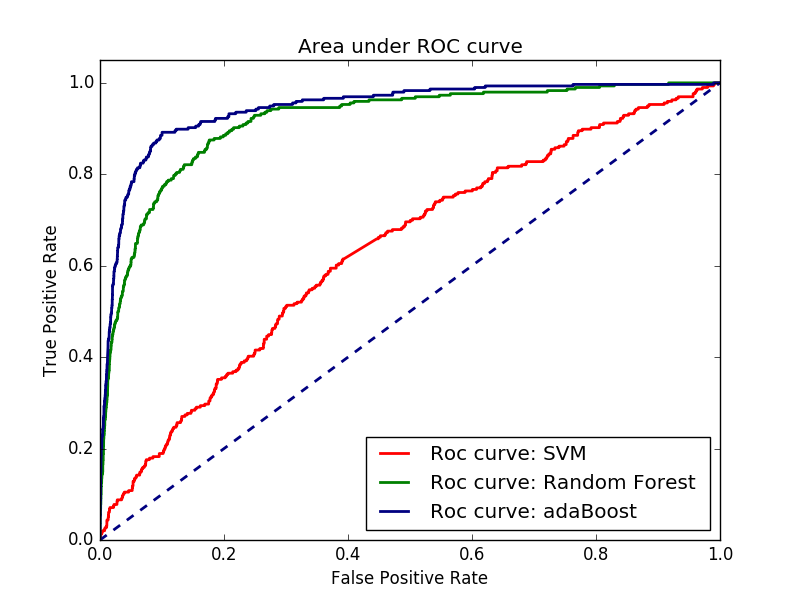} \\
(a)&(b)\\
\includegraphics[width=0.45\textwidth]{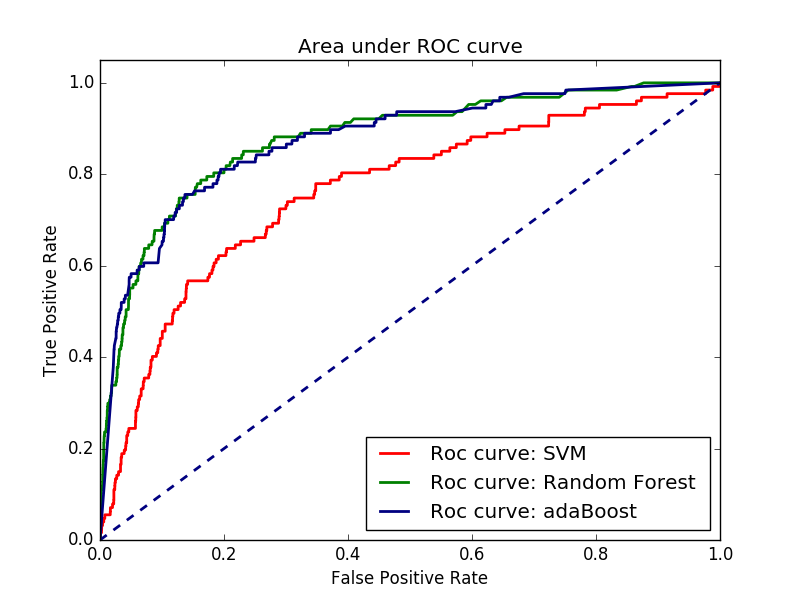}&\includegraphics[width=0.45\textwidth]{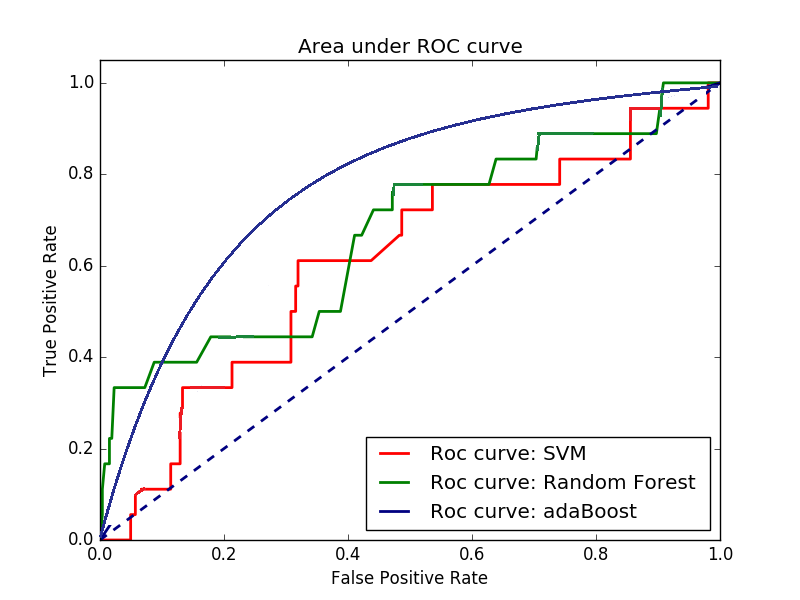} \\
(c)&(d)\\
\end{tabular}
\end{center}
\caption{\csentence{Classifier auROC Analysis.} Receiver operating characteristic curves of different classifier algorithms using random under sampling and all the feature combinations on four datasets: (a) enzymes (b) ion channels (c) GPCRs (d) nuclear receptors.\label{figCLSAUC}}
\end{figure}

In case of the ROC curve, the results are also in support of the selection of AdaBoost as a classifier. AdaBoost provides the highest auROC values for all the four datasets and it gives better auROC values for 11 out of 16 dataset-feature groups combinations. In other cases, SVM has achieved the highest auROC values, but only marginally so. The ROC curves for different classifiers across all feature groups combinations are illustrated in Figure~\ref{figCLSAUC}.

Considering the values of auPR and auROC curves on different datasets as shown in Table~\ref{tabClassifiers} and illustrated through the curves in Figure~\ref{figCLSAUC} and Figure~\ref{figCLSPR}, we select AdaBoost as the classifier for {\methodname}. Note that, because of the huge imbalance in the datasets, with positive samples being much lower than the negative ones, the auPR curve is more important compared to the auROC curve and AdaBoost clearly outperforms the other two classifiers in terms of auPR values.

\subsection*{Effectiveness of the Balancing Methods}
The next set of experiments were run to test the effectiveness of the two different sampling methods on the datasets. The parameters used with AdaBoost classifier for random and cluster based under sampling are reported in Table~\ref{tabParams}.

\begin{figure}[!htb]
\begin{center}
\begin{tabular}{cc}
\includegraphics[width=0.45\textwidth]{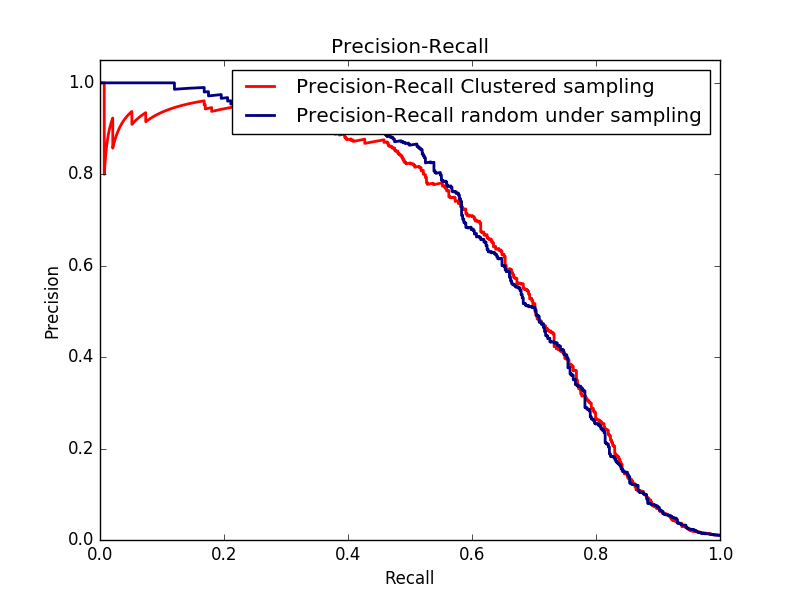}&
\includegraphics[width=0.45\textwidth]{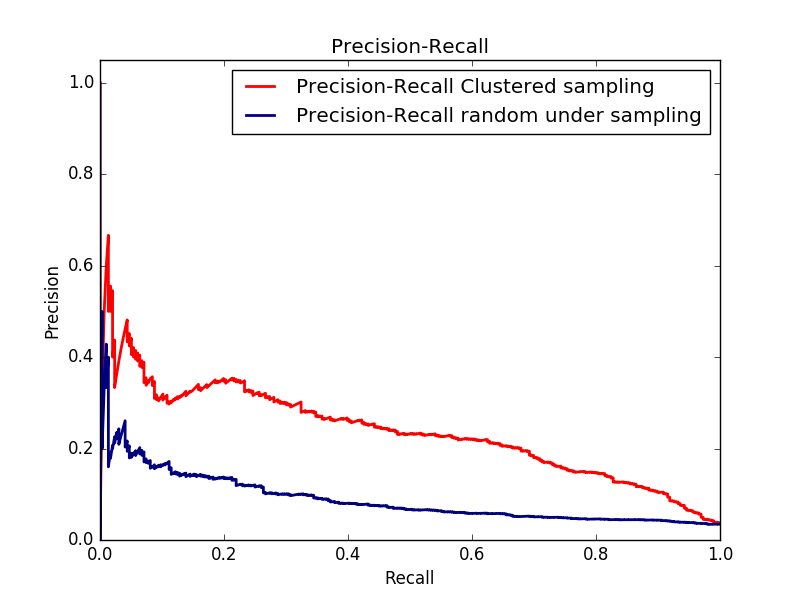} \\
(a)&(b)\\
\includegraphics[width=0.45\textwidth]{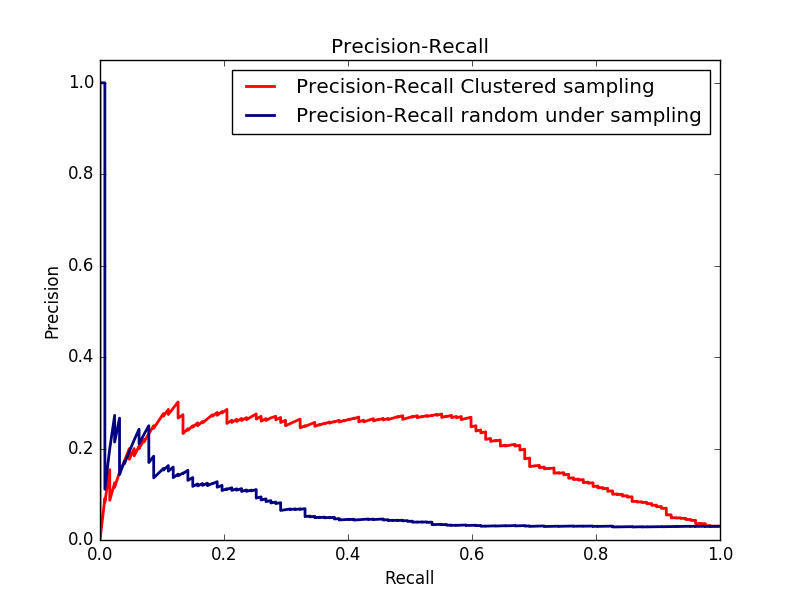}&\includegraphics[width=0.45\textwidth]{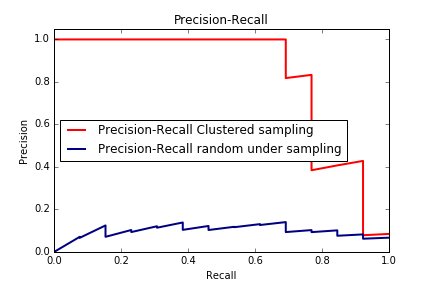} \\
(c)&(d)\\
\end{tabular}
\end{center}
\caption{\csentence{Balancing Method auPR Analysis.} Precision vs Recall curves of AdaBoost classifier showing differences between random under sampling and cluster based sampling using all the feature combinations on four datasets: (a) enzymes (b) ion channels (c) GPCRs (d) nuclear receptors.\label{figBalPR}}
\end{figure}

\begin{figure}[!htb]
\begin{center}
\begin{tabular}{cc}
\includegraphics[width=0.45\textwidth]{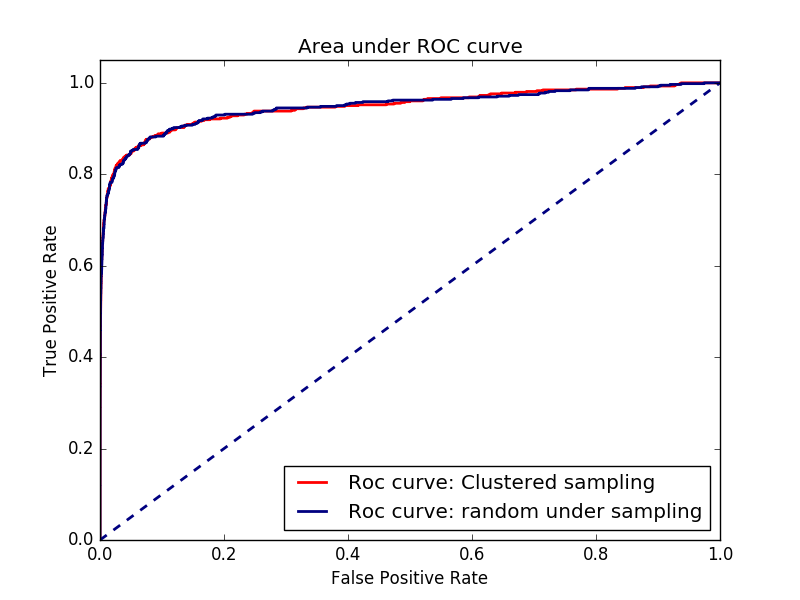}&
\includegraphics[width=0.45\textwidth]{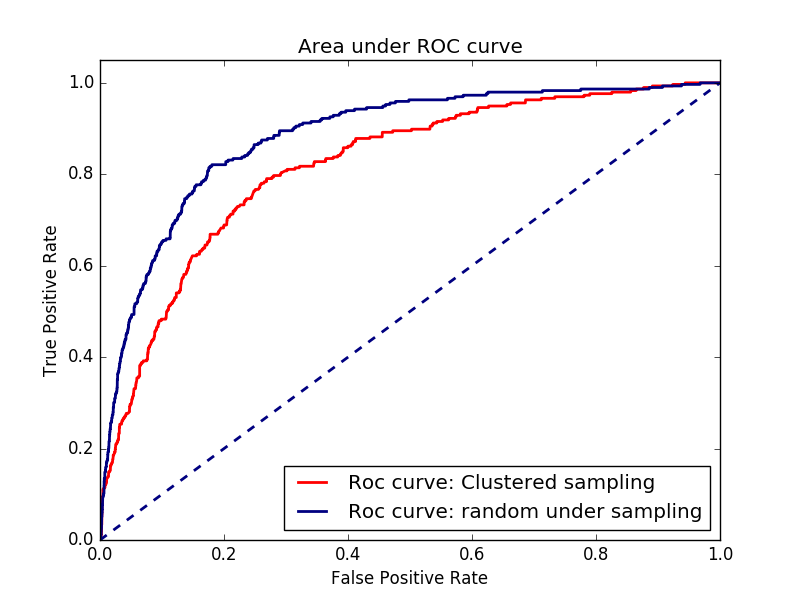} \\
(a)&(b)\\
\includegraphics[width=0.45\textwidth]{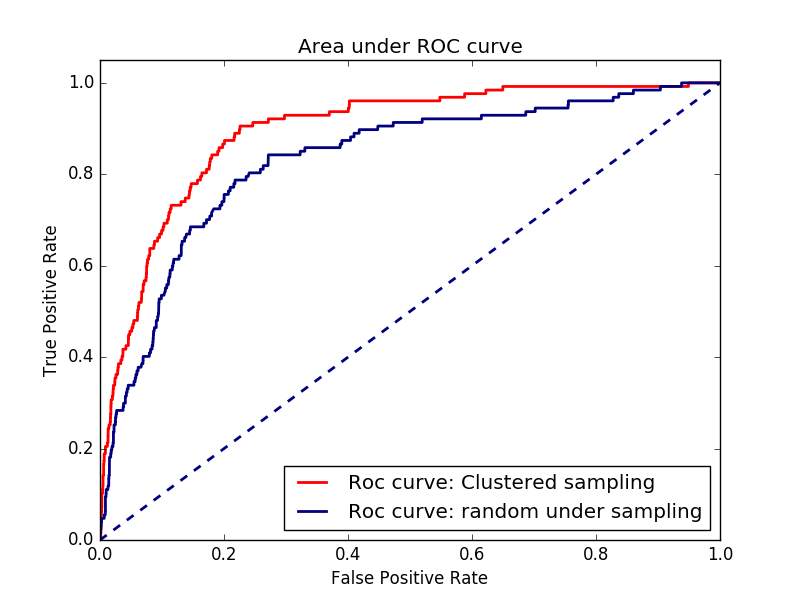}&\includegraphics[width=0.45\textwidth]{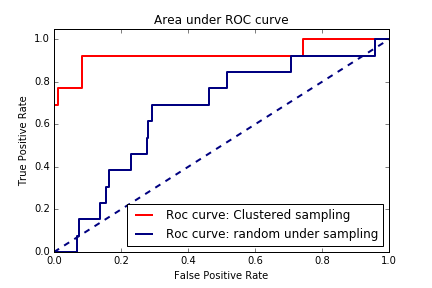} \\
(c)&(d)\\
\end{tabular}

\end{center}
\caption{\csentence{Balancing Method auROC Analysis.} Receiver operating characteristic curves of AdaBoost classifier showing differences between random under sampling and cluster based sampling using all the feature combinations on four datasets: (a) enzymes (b) ion channels (c) GPCRs (d) nuclear receptors.\label{figBalAUC}}
\end{figure}

For each of the datasets, we used four feature group combinations and used random and cluster based under sampling and report auPR and auROC values from cross-validation experiments in Table~\ref{tabFeatBalance}. We also show the ROC curves and auPR curves for all four datasets using all the features in Figure~\ref{figBalAUC} and  Figure~\ref{figBalPR}. %\note{Wrong figure number in one of the two. Please fix.}

From the results reported in Table~\ref{tabFeatBalance}, it is worth-mentioning that in terms of auPR for all four datasets, cluster based sampling significantly outperforms random under sampling method. However, in terms of auROC curve, random sampling is slightly better than cluster based sampling in enzymes and ion channel datasets but the situation is in favor of cluster based sampling in GPCRs and nuclear receptors where it outperforms the random sampling method.

\subsection*{Comparison with Other Methods}
Since the pioneering work of Yamanishi et al. \cite{yamanishi2008prediction}, many supervised learning methods have been applied to predict drug-target interactions on these standard benchmark gold standard datasets. However, a few of these methods \cite{chan2016large,hao2016improved} do not use cross validation techniques and others \cite{keum2017self,wu2017sdtnbi} do not use the same  standard datasets. Our method uses molecular fingerprints and evolutionary and structural features for this supervised classification problem. Similar methods, albeit without  utilizing the structural features and balancing techniques are reported in  \cite{nanni2014set,mousavian2016drug}. Most of the papers in the literature have used auROC curve as the main evaluation metric. We have compared the performance of our method on these four datasets with that of DBSI \cite{cheng2012prediction}, KBMF2K \cite{gonen2012predicting}, NetCBP \cite{chen2013semi}, Yamanishi et al. \cite{yamanishi2008prediction}, Yamanishi et al. \cite{yamanishi2010drug}, Wang et al. \cite{wang2013drug} and Mousavian et al. \cite{mousavian2016drug} using auROC. The auROC values for all these methods along with {\methodname} are reported in Table~\ref{tabMain}.

From the values shown in bold faced font in Table~\ref{tabMain}, we notice that for all the datasets {\methodname} is able to significantly outperform all other previous state-of-the-art methods in terms of auROC. All the auROC values are greater than 90\% which indicates the effectiveness of the classifier, balancing methods and the novel features proposed in this paper.

Moreover, in \cite{mousavian2016drug} the authors argued in favor of auPR curve as a measure of evaluating the performance of classifiers for skewed datasets, especially in drug-target interaction where negative samples outnumber the positive samples. This argument does have merit as, logically, a mis-classification of positive samples or false negative should be more penalized in the score. To compare the performance of our method with that in \cite{mousavian2016drug}, we reported the auPR values of the two predictors in Table~\ref{tabMainPR}. The results clearly shows that our method {\methodname} outperforms the predictor in \cite{mousavian2016drug} in terms of auPR as well.

In Table~\ref{tabOthers}, we report specificity, sensitivity, precision, MCC and F1-Score for four datasets using different feature group combinations as achieved by {\methodname} in experiments. Specificity and sensitivity are very high as reported in this table.

\subsection*{Predicting New Interactions}
In addition to these, we have analyzed the results produced by the classification algorithm. From the false negatives predicted by {\methodname}, we noticed that there are a number of false negatives for which the prediction probability is very high for it to be considered as a negative sample. Similar approaches were adopted in \cite{yamanishi2008prediction,yamanishi2010drug}. In this paper, we suggest that the false negative interactions which are labeled as positive by our method with a very high prediction probability could be potential candidates for finding new positive interactions. A list of such interactions for four group of targets are added as supplementary information with this paper.
	
\subsection*{Web Server Implementation}
We have also implemented our method as shown in Figure~\ref{figTraining} as a separate web server. The web server is freely available for use at: {\website}. The mechanism of the web-server is very simple. We also provide the pre-learned models for each of the datasets. The interface of the web server easy to use. It requires an user first to select the target group and provide the PSSM and SPD files for the target protein. These files can be easily generated by PSI-BLAST and SPIDER2 software using their online available tool.

To specify drug, one can select from a drop down list. The drugs are pre-fetched in our system from KEGG website. After selecting the drug and specifying target files, one can click the prediction button to find the prediction for that drug-target pair. The web-server also have a simple page with easy to-use instructions.

%\note{I will write the conclusion and finalize the abstract. In the mean time, please go through my corrections and if you do not agree with any, revert back or more preferably, discuss with me. PLEASE CHECK ALL FIGURE CAPTIONS CAREFULLY BECAUSE I FOUND SOME TO BE WRONG. As per our telephonic conversation, I still don't find it clear that the testing is done on original situation when balancing is used. Should make it clear at some appropriate point. PLEASE ALSO CHECK MY COMMENTS CAREFULLY AND RESPOND TO THOSE. BTW, a simple command will give us the final version ignoring all comments and track of changes. So, don't worry.} 
\section*{Conclusion}
In this paper, we have presented {\methodname}, a novel method to predict and identify drug-target interactions. {\methodname} is unique in its exploitation of structural features along with the evolutionary features to predict drug-protein interactions. It also uses a novel balancing technique and a boosting technique. We have conducted extensive experiments to test and analyze the performance of {\methodname}. On four benchmark datasets known as the gold standard data in the literature, {\methodname} outperforms the state-of-the-art methods in terms of area under Receiver Operating Characteristic (auROC) curve.

Notably, the gold standard datasets used in the literature as benchmarks to analyze the performance of the methods for drug-target interactions prediction and identification are highly imbalanced with negative samples far outnumbering the positive samples. In the literature it has been argued that area under Precision Recall (auPR) curve is more appropriate as a metric for comparison for such imbalanced datasets. To this end, {\methodname} also outperforms the latest and the best-performing method in the literature to-date in terms of area under precision recall (auPR) curve. We believe that the excellent performance of {\methodname} both in terms of auROC and auPR would motivate the researchers and practitioners to use it to predict drug-target interactions. To facilitate that, {\methodname} is readily available for use at: {\website}.

%%%%%%%%%%%%%%%%%%%%%%%%%%%%%%%%%%%%%%%%%%%%%%
%%                                          %%
%% Backmatter begins here                   %%
%%                                          %%
%%%%%%%%%%%%%%%%%%%%%%%%%%%%%%%%%%%%%%%%%%%%%%

%\begin{backmatter}

\section*{Competing interests}
  The authors declare that they have no competing interests.

\section*{Author's contributions}
    SS initiated the project with the idea of using structural features. FR, SA and DMF equally contributed to the idea of modified balancing method and boosting. FR and SA equally contributed in the implementation and experimentation of the system. All the methods, algorithms and results have been analyzed and verified by SS, DMF, MSR, ZM and AD. All authors contributed significantly in the preparation of the manuscript and approved the final version.

%\section*{Acknowledgements}
 % Text for this section \ldots
%%  After submission of the .TEX file,                     %%
%%  you will be prompted to submit your .BBL file.         %%

% if your bibliography is in bibtex format, use those commands:
\bibliographystyle{bmc-mathphys} % Style BST file (bmc-mathphys, vancouver, spbasic).
\bibliography{drugtarget}      % Bibliography file (usually '*.bib' )


\begin{thebibliography}{64}
\ifx \bisbn   \undefined \def \bisbn  #1{ISBN #1}\fi
\ifx \binits  \undefined \def \binits#1{#1}\fi
\ifx \bauthor  \undefined \def \bauthor#1{#1}\fi
\ifx \batitle  \undefined \def \batitle#1{#1}\fi
\ifx \bjtitle  \undefined \def \bjtitle#1{#1}\fi
\ifx \bvolume  \undefined \def \bvolume#1{\textbf{#1}}\fi
\ifx \byear  \undefined \def \byear#1{#1}\fi
\ifx \bissue  \undefined \def \bissue#1{#1}\fi
\ifx \bfpage  \undefined \def \bfpage#1{#1}\fi
\ifx \blpage  \undefined \def \blpage #1{#1}\fi
\ifx \burl  \undefined \def \burl#1{\textsf{#1}}\fi
\ifx \doiurl  \undefined \def \doiurl#1{\textsf{#1}}\fi
\ifx \betal  \undefined \def \betal{\textit{et al.}}\fi
\ifx \binstitute  \undefined \def \binstitute#1{#1}\fi
\ifx \binstitutionaled  \undefined \def \binstitutionaled#1{#1}\fi
\ifx \bctitle  \undefined \def \bctitle#1{#1}\fi
\ifx \beditor  \undefined \def \beditor#1{#1}\fi
\ifx \bpublisher  \undefined \def \bpublisher#1{#1}\fi
\ifx \bbtitle  \undefined \def \bbtitle#1{#1}\fi
\ifx \bedition  \undefined \def \bedition#1{#1}\fi
\ifx \bseriesno  \undefined \def \bseriesno#1{#1}\fi
\ifx \blocation  \undefined \def \blocation#1{#1}\fi
\ifx \bsertitle  \undefined \def \bsertitle#1{#1}\fi
\ifx \bsnm \undefined \def \bsnm#1{#1}\fi
\ifx \bsuffix \undefined \def \bsuffix#1{#1}\fi
\ifx \bparticle \undefined \def \bparticle#1{#1}\fi
\ifx \barticle \undefined \def \barticle#1{#1}\fi
\ifx \bconfdate \undefined \def \bconfdate #1{#1}\fi
\ifx \botherref \undefined \def \botherref #1{#1}\fi
\ifx \url \undefined \def \url#1{\textsf{#1}}\fi
\ifx \bchapter \undefined \def \bchapter#1{#1}\fi
\ifx \bbook \undefined \def \bbook#1{#1}\fi
\ifx \bcomment \undefined \def \bcomment#1{#1}\fi
\ifx \oauthor \undefined \def \oauthor#1{#1}\fi
\ifx \citeauthoryear \undefined \def \citeauthoryear#1{#1}\fi
\ifx \endbibitem  \undefined \def \endbibitem {}\fi
\ifx \bconflocation  \undefined \def \bconflocation#1{#1}\fi
\ifx \arxivurl  \undefined \def \arxivurl#1{\textsf{#1}}\fi
\csname PreBibitemsHook\endcsname

%%% 1
\bibitem{keiser2009predicting}
\begin{barticle}
\bauthor{\bsnm{Keiser}, \binits{M.J.}},
\bauthor{\bsnm{Setola}, \binits{V.}},
\bauthor{\bsnm{Irwin}, \binits{J.J.}},
\bauthor{\bsnm{Laggner}, \binits{C.}},
\bauthor{\bsnm{Abbas}, \binits{A.I.}},
\bauthor{\bsnm{Hufeisen}, \binits{S.J.}},
\bauthor{\bsnm{Jensen}, \binits{N.H.}},
\bauthor{\bsnm{Kuijer}, \binits{M.B.}},
\bauthor{\bsnm{Matos}, \binits{R.C.}},
\bauthor{\bsnm{Tran}, \binits{T.B.}}, \betal:
\batitle{Predicting new molecular targets for known drugs}.
\bjtitle{Nature}
\bvolume{462}(\bissue{7270}),
\bfpage{175}--\blpage{181}
(\byear{2009})
\end{barticle}
\endbibitem

%%% 2
\bibitem{cheng2013prediction}
\begin{barticle}
\bauthor{\bsnm{Cheng}, \binits{F.}},
\bauthor{\bsnm{Li}, \binits{W.}},
\bauthor{\bsnm{Wu}, \binits{Z.}},
\bauthor{\bsnm{Wang}, \binits{X.}},
\bauthor{\bsnm{Zhang}, \binits{C.}},
\bauthor{\bsnm{Li}, \binits{J.}},
\bauthor{\bsnm{Liu}, \binits{G.}},
\bauthor{\bsnm{Tang}, \binits{Y.}}:
\batitle{Prediction of polypharmacological profiles of drugs by the integration
  of chemical, side effect, and therapeutic space}.
\bjtitle{Journal of chemical information and modeling}
\bvolume{53}(\bissue{4}),
\bfpage{753}--\blpage{762}
(\byear{2013})
\end{barticle}
\endbibitem

%%% 3
\bibitem{wu2017sdtnbi}
\begin{barticle}
\bauthor{\bsnm{Wu}, \binits{Z.}},
\bauthor{\bsnm{Cheng}, \binits{F.}},
\bauthor{\bsnm{Li}, \binits{J.}},
\bauthor{\bsnm{Li}, \binits{W.}},
\bauthor{\bsnm{Liu}, \binits{G.}},
\bauthor{\bsnm{Tang}, \binits{Y.}}:
\batitle{Sdtnbi: an integrated network and chemoinformatics tool for systematic
  prediction of drug--target interactions and drug repositioning}.
\bjtitle{Briefings in bioinformatics}
\bvolume{18}(\bissue{2}),
\bfpage{333}--\blpage{347}
(\byear{2017})
\end{barticle}
\endbibitem

%%% 4
\bibitem{campillos2008drug}
\begin{barticle}
\bauthor{\bsnm{Campillos}, \binits{M.}},
\bauthor{\bsnm{Kuhn}, \binits{M.}},
\bauthor{\bsnm{Gavin}, \binits{A.-C.}},
\bauthor{\bsnm{Jensen}, \binits{L.J.}},
\bauthor{\bsnm{Bork}, \binits{P.}}:
\batitle{Drug target identification using side-effect similarity}.
\bjtitle{Science}
\bvolume{321}(\bissue{5886}),
\bfpage{263}--\blpage{266}
(\byear{2008})
\end{barticle}
\endbibitem

%%% 5
\bibitem{haggarty2003multidimensional}
\begin{barticle}
\bauthor{\bsnm{Haggarty}, \binits{S.J.}},
\bauthor{\bsnm{Koeller}, \binits{K.M.}},
\bauthor{\bsnm{Wong}, \binits{J.C.}},
\bauthor{\bsnm{Butcher}, \binits{R.A.}},
\bauthor{\bsnm{Schreiber}, \binits{S.L.}}:
\batitle{Multidimensional chemical genetic analysis of diversity-oriented
  synthesis-derived deacetylase inhibitors using cell-based assays}.
\bjtitle{Chemistry \& biology}
\bvolume{10}(\bissue{5}),
\bfpage{383}--\blpage{396}
(\byear{2003})
\end{barticle}
\endbibitem

%%% 6
\bibitem{kuruvilla2002dissecting}
\begin{barticle}
\bauthor{\bsnm{Kuruvilla}, \binits{F.G.}},
\bauthor{\bsnm{Shamji}, \binits{A.F.}},
\bauthor{\bsnm{Sternson}, \binits{S.M.}},
\bauthor{\bsnm{Hergenrother}, \binits{P.J.}},
\bauthor{\bsnm{Schreiber}, \binits{S.L.}}:
\batitle{Dissecting glucose signalling with diversity-oriented synthesis and
  small-molecule microarrays}.
\bjtitle{Nature}
\bvolume{416}(\bissue{6881}),
\bfpage{653}--\blpage{657}
(\byear{2002})
\end{barticle}
\endbibitem

%%% 7
\bibitem{hopkins2014role}
\begin{barticle}
\bauthor{\bsnm{Hopkins}, \binits{A.L.}},
\bauthor{\bsnm{Keser{\"u}}, \binits{G.M.}},
\bauthor{\bsnm{Leeson}, \binits{P.D.}},
\bauthor{\bsnm{Rees}, \binits{D.C.}},
\bauthor{\bsnm{Reynolds}, \binits{C.H.}}:
\batitle{The role of ligand efficiency metrics in drug discovery}.
\bjtitle{Nature Reviews Drug Discovery}
\bvolume{13}(\bissue{2}),
\bfpage{105}--\blpage{121}
(\byear{2014})
\end{barticle}
\endbibitem

%%% 8
\bibitem{keiser2007relating}
\begin{barticle}
\bauthor{\bsnm{Keiser}, \binits{M.J.}},
\bauthor{\bsnm{Roth}, \binits{B.L.}},
\bauthor{\bsnm{Armbruster}, \binits{B.N.}},
\bauthor{\bsnm{Ernsberger}, \binits{P.}},
\bauthor{\bsnm{Irwin}, \binits{J.J.}},
\bauthor{\bsnm{Shoichet}, \binits{B.K.}}:
\batitle{Relating protein pharmacology by ligand chemistry}.
\bjtitle{Nature biotechnology}
\bvolume{25}(\bissue{2}),
\bfpage{197}--\blpage{206}
(\byear{2007})
\end{barticle}
\endbibitem

%%% 9
\bibitem{ma2013drug}
\begin{barticle}
\bauthor{\bsnm{Ma}, \binits{D.-L.}},
\bauthor{\bsnm{Chan}, \binits{D.S.-H.}},
\bauthor{\bsnm{Leung}, \binits{C.-H.}}:
\batitle{Drug repositioning by structure-based virtual screening}.
\bjtitle{Chemical Society Reviews}
\bvolume{42}(\bissue{5}),
\bfpage{2130}--\blpage{2141}
(\byear{2013})
\end{barticle}
\endbibitem

%%% 10
\bibitem{pan2013molecular}
\begin{barticle}
\bauthor{\bsnm{Pan}, \binits{A.C.}},
\bauthor{\bsnm{Borhani}, \binits{D.W.}},
\bauthor{\bsnm{Dror}, \binits{R.O.}},
\bauthor{\bsnm{Shaw}, \binits{D.E.}}:
\batitle{Molecular determinants of drug--receptor binding kinetics}.
\bjtitle{Drug discovery today}
\bvolume{18}(\bissue{13}),
\bfpage{667}--\blpage{673}
(\byear{2013})
\end{barticle}
\endbibitem

%%% 11
\bibitem{mutowo2016drug}
\begin{barticle}
\bauthor{\bsnm{Mutowo}, \binits{P.}},
\bauthor{\bsnm{Bento}, \binits{A.P.}},
\bauthor{\bsnm{Dedman}, \binits{N.}},
\bauthor{\bsnm{Gaulton}, \binits{A.}},
\bauthor{\bsnm{Hersey}, \binits{A.}},
\bauthor{\bsnm{Lomax}, \binits{J.}},
\bauthor{\bsnm{Overington}, \binits{J.P.}}:
\batitle{A drug target slim: using gene ontology and gene ontology annotations
  to navigate protein-ligand target space in chembl}.
\bjtitle{Journal of biomedical semantics}
\bvolume{7}(\bissue{1}),
\bfpage{59}
(\byear{2016})
\end{barticle}
\endbibitem

%%% 12
\bibitem{plake2011computational}
\begin{barticle}
\bauthor{\bsnm{Plake}, \binits{C.}},
\bauthor{\bsnm{Schroeder}, \binits{M.}}:
\batitle{Computational polypharmacology with text mining and ontologies}.
\bjtitle{Current pharmaceutical biotechnology}
\bvolume{12}(\bissue{3}),
\bfpage{449}--\blpage{457}
(\byear{2011})
\end{barticle}
\endbibitem

%%% 13
\bibitem{zhu2005probabilistic}
\begin{barticle}
\bauthor{\bsnm{Zhu}, \binits{S.}},
\bauthor{\bsnm{Okuno}, \binits{Y.}},
\bauthor{\bsnm{Tsujimoto}, \binits{G.}},
\bauthor{\bsnm{Mamitsuka}, \binits{H.}}:
\batitle{A probabilistic model for mining implicit ‘chemical
  compound--gene’relations from literature}.
\bjtitle{Bioinformatics}
\bvolume{21}(\bissue{suppl 2}),
\bfpage{245}--\blpage{251}
(\byear{2005})
\end{barticle}
\endbibitem

%%% 14
\bibitem{morris2009autodock4}
\begin{barticle}
\bauthor{\bsnm{Morris}, \binits{G.M.}},
\bauthor{\bsnm{Huey}, \binits{R.}},
\bauthor{\bsnm{Lindstrom}, \binits{W.}},
\bauthor{\bsnm{Sanner}, \binits{M.F.}},
\bauthor{\bsnm{Belew}, \binits{R.K.}},
\bauthor{\bsnm{Goodsell}, \binits{D.S.}},
\bauthor{\bsnm{Olson}, \binits{A.J.}}:
\batitle{Autodock4 and autodocktools4: Automated docking with selective
  receptor flexibility}.
\bjtitle{Journal of computational chemistry}
\bvolume{30}(\bissue{16}),
\bfpage{2785}--\blpage{2791}
(\byear{2009})
\end{barticle}
\endbibitem

%%% 15
\bibitem{mousavian2014drug}
\begin{barticle}
\bauthor{\bsnm{Mousavian}, \binits{Z.}},
\bauthor{\bsnm{Masoudi-Nejad}, \binits{A.}}:
\batitle{Drug--target interaction prediction via chemogenomic space:
  learning-based methods}.
\bjtitle{Expert opinion on drug metabolism \& toxicology}
\bvolume{10}(\bissue{9}),
\bfpage{1273}--\blpage{1287}
(\byear{2014})
\end{barticle}
\endbibitem

%%% 16
\bibitem{yamanishi2008prediction}
\begin{barticle}
\bauthor{\bsnm{Yamanishi}, \binits{Y.}},
\bauthor{\bsnm{Araki}, \binits{M.}},
\bauthor{\bsnm{Gutteridge}, \binits{A.}},
\bauthor{\bsnm{Honda}, \binits{W.}},
\bauthor{\bsnm{Kanehisa}, \binits{M.}}:
\batitle{Prediction of drug--target interaction networks from the integration
  of chemical and genomic spaces}.
\bjtitle{Bioinformatics}
\bvolume{24}(\bissue{13}),
\bfpage{232}--\blpage{240}
(\byear{2008})
\end{barticle}
\endbibitem

%%% 17
\bibitem{bleakley2009supervised}
\begin{barticle}
\bauthor{\bsnm{Bleakley}, \binits{K.}},
\bauthor{\bsnm{Yamanishi}, \binits{Y.}}:
\batitle{Supervised prediction of drug--target interactions using bipartite
  local models}.
\bjtitle{Bioinformatics}
\bvolume{25}(\bissue{18}),
\bfpage{2397}--\blpage{2403}
(\byear{2009})
\end{barticle}
\endbibitem

%%% 18
\bibitem{wang2013drug}
\begin{bchapter}
\bauthor{\bsnm{Wang}, \binits{W.}},
\bauthor{\bsnm{Yang}, \binits{S.}},
\bauthor{\bsnm{Li}, \binits{J.}}:
\bctitle{Drug target predictions based on heterogeneous graph inference}.
In: \bbtitle{Pacific Symposium on Biocomputing. Pacific Symposium on
  Biocomputing},
p. \bfpage{53}
(\byear{2013}).
\bcomment{NIH Public Access}
\end{bchapter}
\endbibitem

%%% 19
\bibitem{chen2012drug}
\begin{barticle}
\bauthor{\bsnm{Chen}, \binits{X.}},
\bauthor{\bsnm{Liu}, \binits{M.-X.}},
\bauthor{\bsnm{Yan}, \binits{G.-Y.}}:
\batitle{Drug--target interaction prediction by random walk on the
  heterogeneous network}.
\bjtitle{Molecular BioSystems}
\bvolume{8}(\bissue{7}),
\bfpage{1970}--\blpage{1978}
(\byear{2012})
\end{barticle}
\endbibitem

%%% 20
\bibitem{alaimo2013drug}
\begin{barticle}
\bauthor{\bsnm{Alaimo}, \binits{S.}},
\bauthor{\bsnm{Pulvirenti}, \binits{A.}},
\bauthor{\bsnm{Giugno}, \binits{R.}},
\bauthor{\bsnm{Ferro}, \binits{A.}}:
\batitle{Drug--target interaction prediction through domain-tuned network-based
  inference}.
\bjtitle{Bioinformatics}
\bvolume{29}(\bissue{16}),
\bfpage{2004}--\blpage{2008}
(\byear{2013})
\end{barticle}
\endbibitem

%%% 21
\bibitem{cheng2012prediction}
\begin{barticle}
\bauthor{\bsnm{Cheng}, \binits{F.}},
\bauthor{\bsnm{Liu}, \binits{C.}},
\bauthor{\bsnm{Jiang}, \binits{J.}},
\bauthor{\bsnm{Lu}, \binits{W.}},
\bauthor{\bsnm{Li}, \binits{W.}},
\bauthor{\bsnm{Liu}, \binits{G.}},
\bauthor{\bsnm{Zhou}, \binits{W.}},
\bauthor{\bsnm{Huang}, \binits{J.}},
\bauthor{\bsnm{Tang}, \binits{Y.}}:
\batitle{Prediction of drug-target interactions and drug repositioning via
  network-based inference}.
\bjtitle{PLoS Comput Biol}
\bvolume{8}(\bissue{5}),
\bfpage{1002503}
(\byear{2012})
\end{barticle}
\endbibitem

%%% 22
\bibitem{mousavian2016drug}
\begin{barticle}
\bauthor{\bsnm{Mousavian}, \binits{Z.}},
\bauthor{\bsnm{Khakabimamaghani}, \binits{S.}},
\bauthor{\bsnm{Kavousi}, \binits{K.}},
\bauthor{\bsnm{Masoudi-Nejad}, \binits{A.}}:
\batitle{Drug--target interaction prediction from pssm based evolutionary
  information}.
\bjtitle{Journal of pharmacological and toxicological methods}
\bvolume{78},
\bfpage{42}--\blpage{51}
(\byear{2016})
\end{barticle}
\endbibitem

%%% 23
\bibitem{keum2017self}
\begin{barticle}
\bauthor{\bsnm{Keum}, \binits{J.}},
\bauthor{\bsnm{Nam}, \binits{H.}}:
\batitle{Self-blm: Prediction of drug-target interactions via self-training
  svm}.
\bjtitle{PloS one}
\bvolume{12}(\bissue{2}),
\bfpage{0171839}
(\byear{2017})
\end{barticle}
\endbibitem

%%% 24
\bibitem{chan2016large}
\begin{bchapter}
\bauthor{\bsnm{Chan}, \binits{K.C.}},
\bauthor{\bsnm{You}, \binits{Z.-H.}}, \betal:
\bctitle{Large-scale prediction of drug-target interactions from deep
  representations}.
In: \bbtitle{Neural Networks (IJCNN), 2016 International Joint Conference On},
pp. \bfpage{1236}--\blpage{1243}
(\byear{2016}).
\bcomment{IEEE}
\end{bchapter}
\endbibitem

%%% 25
\bibitem{xiao2013icdi}
\begin{barticle}
\bauthor{\bsnm{Xiao}, \binits{X.}},
\bauthor{\bsnm{Min}, \binits{J.-L.}},
\bauthor{\bsnm{Wang}, \binits{P.}},
\bauthor{\bsnm{Chou}, \binits{K.-C.}}:
\batitle{icdi-psefpt: identify the channel--drug interaction in cellular
  networking with pseaac and molecular fingerprints}.
\bjtitle{Journal of theoretical biology}
\bvolume{337},
\bfpage{71}--\blpage{79}
(\byear{2013})
\end{barticle}
\endbibitem

%%% 26
\bibitem{he2010predicting}
\begin{barticle}
\bauthor{\bsnm{He}, \binits{Z.}},
\bauthor{\bsnm{Zhang}, \binits{J.}},
\bauthor{\bsnm{Shi}, \binits{X.-H.}},
\bauthor{\bsnm{Hu}, \binits{L.-L.}},
\bauthor{\bsnm{Kong}, \binits{X.}},
\bauthor{\bsnm{Cai}, \binits{Y.-D.}},
\bauthor{\bsnm{Chou}, \binits{K.-C.}}:
\batitle{Predicting drug-target interaction networks based on functional groups
  and biological features}.
\bjtitle{PloS one}
\bvolume{5}(\bissue{3}),
\bfpage{9603}
(\byear{2010})
\end{barticle}
\endbibitem

%%% 27
\bibitem{yamanishi2010drug}
\begin{barticle}
\bauthor{\bsnm{Yamanishi}, \binits{Y.}},
\bauthor{\bsnm{Kotera}, \binits{M.}},
\bauthor{\bsnm{Kanehisa}, \binits{M.}},
\bauthor{\bsnm{Goto}, \binits{S.}}:
\batitle{Drug-target interaction prediction from chemical, genomic and
  pharmacological data in an integrated framework}.
\bjtitle{Bioinformatics}
\bvolume{26}(\bissue{12}),
\bfpage{246}--\blpage{254}
(\byear{2010})
\end{barticle}
\endbibitem

%%% 28
\bibitem{hao2016improved}
\begin{barticle}
\bauthor{\bsnm{Hao}, \binits{M.}},
\bauthor{\bsnm{Wang}, \binits{Y.}},
\bauthor{\bsnm{Bryant}, \binits{S.H.}}:
\batitle{Improved prediction of drug-target interactions using regularized
  least squares integrating with kernel fusion technique}.
\bjtitle{Analytica chimica acta}
\bvolume{909},
\bfpage{41}--\blpage{50}
(\byear{2016})
\end{barticle}
\endbibitem

%%% 29
\bibitem{gonen2012predicting}
\begin{barticle}
\bauthor{\bsnm{G{\"o}nen}, \binits{M.}}:
\batitle{Predicting drug--target interactions from chemical and genomic kernels
  using bayesian matrix factorization}.
\bjtitle{Bioinformatics}
\bvolume{28}(\bissue{18}),
\bfpage{2304}--\blpage{2310}
(\byear{2012})
\end{barticle}
\endbibitem

%%% 30
\bibitem{chen2013semi}
\begin{barticle}
\bauthor{\bsnm{Chen}, \binits{H.}},
\bauthor{\bsnm{Zhang}, \binits{Z.}}:
\batitle{A semi-supervised method for drug-target interaction prediction with
  consistency in networks}.
\bjtitle{PloS one}
\bvolume{8}(\bissue{5}),
\bfpage{62975}
(\byear{2013})
\end{barticle}
\endbibitem

%%% 31
\bibitem{ba2016daspfind}
\begin{barticle}
\bauthor{\bsnm{Ba-Alawi}, \binits{W.}},
\bauthor{\bsnm{Soufan}, \binits{O.}},
\bauthor{\bsnm{Essack}, \binits{M.}},
\bauthor{\bsnm{Kalnis}, \binits{P.}},
\bauthor{\bsnm{Bajic}, \binits{V.B.}}:
\batitle{Daspfind: new efficient method to predict drug--target interactions}.
\bjtitle{Journal of cheminformatics}
\bvolume{8}(\bissue{1}),
\bfpage{15}
(\byear{2016})
\end{barticle}
\endbibitem

%%% 32
\bibitem{yang2016sixty}
\begin{botherref}
\oauthor{\bsnm{Yang}, \binits{Y.}},
\oauthor{\bsnm{Gao}, \binits{J.}},
\oauthor{\bsnm{Wang}, \binits{J.}},
\oauthor{\bsnm{Heffernan}, \binits{R.}},
\oauthor{\bsnm{Hanson}, \binits{J.}},
\oauthor{\bsnm{Paliwal}, \binits{K.}},
\oauthor{\bsnm{Zhou}, \binits{Y.}}:
Sixty-five years of the long march in protein secondary structure prediction:
  the final stretch?
Briefings in Bioinformatics,
129
(2016)
\end{botherref}
\endbibitem

%%% 33
\bibitem{yang2017spider2}
\begin{botherref}
\oauthor{\bsnm{Yang}, \binits{Y.}},
\oauthor{\bsnm{Heffernan}, \binits{R.}},
\oauthor{\bsnm{Paliwal}, \binits{K.}},
\oauthor{\bsnm{Lyons}, \binits{J.}},
\oauthor{\bsnm{Dehzangi}, \binits{A.}},
\oauthor{\bsnm{Sharma}, \binits{A.}},
\oauthor{\bsnm{Wang}, \binits{J.}},
\oauthor{\bsnm{Sattar}, \binits{A.}},
\oauthor{\bsnm{Zhou}, \binits{Y.}}:
Spider2: A package to predict secondary structure, accessible surface area, and
  main-chain torsional angles by deep neural networks.
Prediction of Protein Secondary Structure,
55--63
(2017)
\end{botherref}
\endbibitem

%%% 34
\bibitem{lopez2017sucstruct}
\begin{botherref}
\oauthor{\bsnm{L{\'o}pez}, \binits{Y.}},
\oauthor{\bsnm{Dehzangi}, \binits{A.}},
\oauthor{\bsnm{Lal}, \binits{S.P.}},
\oauthor{\bsnm{Taherzadeh}, \binits{G.}},
\oauthor{\bsnm{Michaelson}, \binits{J.}},
\oauthor{\bsnm{Sattar}, \binits{A.}},
\oauthor{\bsnm{Tsunoda}, \binits{T.}},
\oauthor{\bsnm{Sharma}, \binits{A.}}:
Sucstruct: Prediction of succinylated lysine residues by using structural
  properties of amino acids.
Analytical Biochemistry
(2017)
\end{botherref}
\endbibitem

%%% 35
\bibitem{chou2011some}
\begin{barticle}
\bauthor{\bsnm{Chou}, \binits{K.-C.}}:
\batitle{Some remarks on protein attribute prediction and pseudo amino acid
  composition}.
\bjtitle{Journal of theoretical biology}
\bvolume{273}(\bissue{1}),
\bfpage{236}--\blpage{247}
(\byear{2011})
\end{barticle}
\endbibitem

%%% 36
\bibitem{knox2011drugbank}
\begin{barticle}
\bauthor{\bsnm{Knox}, \binits{C.}},
\bauthor{\bsnm{Law}, \binits{V.}},
\bauthor{\bsnm{Jewison}, \binits{T.}},
\bauthor{\bsnm{Liu}, \binits{P.}},
\bauthor{\bsnm{Ly}, \binits{S.}},
\bauthor{\bsnm{Frolkis}, \binits{A.}},
\bauthor{\bsnm{Pon}, \binits{A.}},
\bauthor{\bsnm{Banco}, \binits{K.}},
\bauthor{\bsnm{Mak}, \binits{C.}},
\bauthor{\bsnm{Neveu}, \binits{V.}}, \betal:
\batitle{Drugbank 3.0: a comprehensive resource for ‘omics’ research on
  drugs}.
\bjtitle{Nucleic acids research}
\bvolume{39}(\bissue{suppl 1}),
\bfpage{1035}--\blpage{1041}
(\byear{2011})
\end{barticle}
\endbibitem

%%% 37
\bibitem{kanehisa2000kegg}
\begin{barticle}
\bauthor{\bsnm{Kanehisa}, \binits{M.}},
\bauthor{\bsnm{Goto}, \binits{S.}}:
\batitle{Kegg: kyoto encyclopedia of genes and genomes}.
\bjtitle{Nucleic acids research}
\bvolume{28}(\bissue{1}),
\bfpage{27}--\blpage{30}
(\byear{2000})
\end{barticle}
\endbibitem

%%% 38
\bibitem{altschul1997gapped}
\begin{barticle}
\bauthor{\bsnm{Altschul}, \binits{S.F.}},
\bauthor{\bsnm{Madden}, \binits{T.L.}},
\bauthor{\bsnm{Sch{\"a}ffer}, \binits{A.A.}},
\bauthor{\bsnm{Zhang}, \binits{J.}},
\bauthor{\bsnm{Zhang}, \binits{Z.}},
\bauthor{\bsnm{Miller}, \binits{W.}},
\bauthor{\bsnm{Lipman}, \binits{D.J.}}:
\batitle{Gapped blast and psi-blast: a new generation of protein database
  search programs}.
\bjtitle{Nucleic acids research}
\bvolume{25}(\bissue{17}),
\bfpage{3389}--\blpage{3402}
(\byear{1997})
\end{barticle}
\endbibitem

%%% 39
\bibitem{wishart2008drugbank}
\begin{barticle}
\bauthor{\bsnm{Wishart}, \binits{D.S.}},
\bauthor{\bsnm{Knox}, \binits{C.}},
\bauthor{\bsnm{Guo}, \binits{A.C.}},
\bauthor{\bsnm{Cheng}, \binits{D.}},
\bauthor{\bsnm{Shrivastava}, \binits{S.}},
\bauthor{\bsnm{Tzur}, \binits{D.}},
\bauthor{\bsnm{Gautam}, \binits{B.}},
\bauthor{\bsnm{Hassanali}, \binits{M.}}:
\batitle{Drugbank: a knowledgebase for drugs, drug actions and drug targets}.
\bjtitle{Nucleic acids research}
\bvolume{36}(\bissue{suppl 1}),
\bfpage{901}--\blpage{906}
(\byear{2008})
\end{barticle}
\endbibitem

%%% 40
\bibitem{kanehisa2008kegg}
\begin{barticle}
\bauthor{\bsnm{Kanehisa}, \binits{M.}},
\bauthor{\bsnm{Araki}, \binits{M.}},
\bauthor{\bsnm{Goto}, \binits{S.}},
\bauthor{\bsnm{Hattori}, \binits{M.}},
\bauthor{\bsnm{Hirakawa}, \binits{M.}},
\bauthor{\bsnm{Itoh}, \binits{M.}},
\bauthor{\bsnm{Katayama}, \binits{T.}},
\bauthor{\bsnm{Kawashima}, \binits{S.}},
\bauthor{\bsnm{Okuda}, \binits{S.}},
\bauthor{\bsnm{Tokimatsu}, \binits{T.}}, \betal:
\batitle{Kegg for linking genomes to life and the environment}.
\bjtitle{Nucleic acids research}
\bvolume{36}(\bissue{suppl 1}),
\bfpage{480}--\blpage{484}
(\byear{2008})
\end{barticle}
\endbibitem

%%% 41
\bibitem{schomburg2004brenda}
\begin{barticle}
\bauthor{\bsnm{Schomburg}, \binits{I.}},
\bauthor{\bsnm{Chang}, \binits{A.}},
\bauthor{\bsnm{Ebeling}, \binits{C.}},
\bauthor{\bsnm{Gremse}, \binits{M.}},
\bauthor{\bsnm{Heldt}, \binits{C.}},
\bauthor{\bsnm{Huhn}, \binits{G.}},
\bauthor{\bsnm{Schomburg}, \binits{D.}}:
\batitle{Brenda, the enzyme database: updates and major new developments}.
\bjtitle{Nucleic acids research}
\bvolume{32}(\bissue{suppl 1}),
\bfpage{431}--\blpage{433}
(\byear{2004})
\end{barticle}
\endbibitem

%%% 42
\bibitem{gunther2008supertarget}
\begin{barticle}
\bauthor{\bsnm{G{\"u}nther}, \binits{S.}},
\bauthor{\bsnm{Kuhn}, \binits{M.}},
\bauthor{\bsnm{Dunkel}, \binits{M.}},
\bauthor{\bsnm{Campillos}, \binits{M.}},
\bauthor{\bsnm{Senger}, \binits{C.}},
\bauthor{\bsnm{Petsalaki}, \binits{E.}},
\bauthor{\bsnm{Ahmed}, \binits{J.}},
\bauthor{\bsnm{Urdiales}, \binits{E.G.}},
\bauthor{\bsnm{Gewiess}, \binits{A.}},
\bauthor{\bsnm{Jensen}, \binits{L.J.}}, \betal:
\batitle{Supertarget and matador: resources for exploring drug-target
  relationships}.
\bjtitle{Nucleic acids research}
\bvolume{36}(\bissue{suppl 1}),
\bfpage{919}--\blpage{922}
(\byear{2008})
\end{barticle}
\endbibitem

%%% 43
\bibitem{todeschini2008handbook}
\begin{bbook}
\bauthor{\bsnm{Todeschini}, \binits{R.}},
\bauthor{\bsnm{Consonni}, \binits{V.}}:
\bbtitle{Handbook of Molecular Descriptors}
vol. \bseriesno{11}.
\bpublisher{John Wiley \& Sons}, \blocation{???}
(\byear{2008})
\end{bbook}
\endbibitem

%%% 44
\bibitem{tabei2013scalable}
\begin{barticle}
\bauthor{\bsnm{Tabei}, \binits{Y.}},
\bauthor{\bsnm{Yamanishi}, \binits{Y.}}:
\batitle{Scalable prediction of compound-protein interactions using minwise
  hashing}.
\bjtitle{BMC systems biology}
\bvolume{7}(\bissue{6}),
\bfpage{3}
(\byear{2013})
\end{barticle}
\endbibitem

%%% 45
\bibitem{tabei2012identification}
\begin{barticle}
\bauthor{\bsnm{Tabei}, \binits{Y.}},
\bauthor{\bsnm{Pauwels}, \binits{E.}},
\bauthor{\bsnm{Stoven}, \binits{V.}},
\bauthor{\bsnm{Takemoto}, \binits{K.}},
\bauthor{\bsnm{Yamanishi}, \binits{Y.}}:
\batitle{Identification of chemogenomic features from drug--target interaction
  networks using interpretable classifiers}.
\bjtitle{Bioinformatics}
\bvolume{28}(\bissue{18}),
\bfpage{487}--\blpage{494}
(\byear{2012})
\end{barticle}
\endbibitem

%%% 46
\bibitem{chen2009pubchem}
\begin{barticle}
\bauthor{\bsnm{Chen}, \binits{B.}},
\bauthor{\bsnm{Wild}, \binits{D.}},
\bauthor{\bsnm{Guha}, \binits{R.}}:
\batitle{Pubchem as a source of polypharmacology}.
\bjtitle{Journal of chemical information and modeling}
\bvolume{49}(\bissue{9}),
\bfpage{2044}--\blpage{2055}
(\byear{2009})
\end{barticle}
\endbibitem

%%% 47
\bibitem{guha2007chemical}
\begin{barticle}
\bauthor{\bsnm{Guha}, \binits{R.}}, \betal:
\batitle{Chemical informatics functionality in r}.
\bjtitle{Journal of Statistical Software}
\bvolume{18}(\bissue{5}),
\bfpage{1}--\blpage{16}
(\byear{2007})
\end{barticle}
\endbibitem

%%% 48
\bibitem{sharma2015predict}
\begin{barticle}
\bauthor{\bsnm{Sharma}, \binits{R.}},
\bauthor{\bsnm{Dehzangi}, \binits{A.}},
\bauthor{\bsnm{Lyons}, \binits{J.}},
\bauthor{\bsnm{Paliwal}, \binits{K.}},
\bauthor{\bsnm{Tsunoda}, \binits{T.}},
\bauthor{\bsnm{Sharma}, \binits{A.}}:
\batitle{Predict gram-positive and gram-negative subcellular localization via
  incorporating evolutionary information and physicochemical features into
  chou's general pseaac}.
\bjtitle{IEEE Transactions on NanoBioscience}
\bvolume{14}(\bissue{8}),
\bfpage{915}--\blpage{926}
(\byear{2015})
\end{barticle}
\endbibitem

%%% 49
\bibitem{paliwal2014tri}
\begin{barticle}
\bauthor{\bsnm{Paliwal}, \binits{K.K.}},
\bauthor{\bsnm{Sharma}, \binits{A.}},
\bauthor{\bsnm{Lyons}, \binits{J.}},
\bauthor{\bsnm{Dehzangi}, \binits{A.}}:
\batitle{A tri-gram based feature extraction technique using linear
  probabilities of position specific scoring matrix for protein fold
  recognition}.
\bjtitle{IEEE Transactions on Nanobioscience}
\bvolume{13}(\bissue{1}),
\bfpage{44}--\blpage{50}
(\byear{2014})
\end{barticle}
\endbibitem

%%% 50
\bibitem{chawla2002smote}
\begin{barticle}
\bauthor{\bsnm{Chawla}, \binits{N.V.}},
\bauthor{\bsnm{Bowyer}, \binits{K.W.}},
\bauthor{\bsnm{Hall}, \binits{L.O.}},
\bauthor{\bsnm{Kegelmeyer}, \binits{W.P.}}:
\batitle{Smote: synthetic minority over-sampling technique}.
\bjtitle{Journal of artificial intelligence research}
\bvolume{16},
\bfpage{321}--\blpage{357}
(\byear{2002})
\end{barticle}
\endbibitem

%%% 51
\bibitem{yu2010simple}
\begin{barticle}
\bauthor{\bsnm{Yu}, \binits{J.}},
\bauthor{\bsnm{Guo}, \binits{M.}},
\bauthor{\bsnm{Needham}, \binits{C.J.}},
\bauthor{\bsnm{Huang}, \binits{Y.}},
\bauthor{\bsnm{Cai}, \binits{L.}},
\bauthor{\bsnm{Westhead}, \binits{D.R.}}:
\batitle{Simple sequence-based kernels do not predict protein--protein
  interactions}.
\bjtitle{Bioinformatics}
\bvolume{26}(\bissue{20}),
\bfpage{2610}--\blpage{2614}
(\byear{2010})
\end{barticle}
\endbibitem

%%% 52
\bibitem{laurikkala2001improving}
\begin{bchapter}
\bauthor{\bsnm{Laurikkala}, \binits{J.}}:
\bctitle{Improving identification of difficult small classes by balancing class
  distribution}.
In: \bbtitle{Conference on Artificial Intelligence in Medicine in Europe},
pp. \bfpage{63}--\blpage{66}
(\byear{2001}).
\bcomment{Springer}
\end{bchapter}
\endbibitem

%%% 53
\bibitem{yen2009cluster}
\begin{barticle}
\bauthor{\bsnm{Yen}, \binits{S.-J.}},
\bauthor{\bsnm{Lee}, \binits{Y.-S.}}:
\batitle{Cluster-based under-sampling approaches for imbalanced data
  distributions}.
\bjtitle{Expert Systems with Applications}
\bvolume{36}(\bissue{3}),
\bfpage{5718}--\blpage{5727}
(\byear{2009})
\end{barticle}
\endbibitem

%%% 54
\bibitem{rahman2013cluster}
\begin{bchapter}
\bauthor{\bsnm{Rahman}, \binits{M.M.}},
\bauthor{\bsnm{Davis}, \binits{D.}}:
\bctitle{Cluster based under-sampling for unbalanced cardiovascular data}.
In: \bbtitle{Proceedings of the World Congress on Engineering},
vol. \bseriesno{3},
pp. \bfpage{3}--\blpage{5}
(\byear{2013})
\end{bchapter}
\endbibitem

%%% 55
\bibitem{freund1995desicion}
\begin{bchapter}
\bauthor{\bsnm{Freund}, \binits{Y.}},
\bauthor{\bsnm{Schapire}, \binits{R.E.}}:
\bctitle{A desicion-theoretic generalization of on-line learning and an
  application to boosting}.
In: \bbtitle{European Conference on Computational Learning Theory},
pp. \bfpage{23}--\blpage{37}
(\byear{1995}).
\bcomment{Springer}
\end{bchapter}
\endbibitem

%%% 56
\bibitem{mohri2012foundations}
\begin{bbook}
\bauthor{\bsnm{Mohri}, \binits{M.}},
\bauthor{\bsnm{Rostamizadeh}, \binits{A.}},
\bauthor{\bsnm{Talwalkar}, \binits{A.}}:
\bbtitle{Foundations of Machine Learning}.
\bpublisher{MIT press}, \blocation{???}
(\byear{2012})
\end{bbook}
\endbibitem

%%% 57
\bibitem{powers2011evaluation}
\begin{botherref}
\oauthor{\bsnm{Powers}, \binits{D.M.}}:
Evaluation: from precision, recall and f-measure to roc, informedness,
  markedness and correlation
(2011)
\end{botherref}
\endbibitem

%%% 58
\bibitem{cao2012large}
\begin{barticle}
\bauthor{\bsnm{Cao}, \binits{D.-S.}},
\bauthor{\bsnm{Liu}, \binits{S.}},
\bauthor{\bsnm{Xu}, \binits{Q.-S.}},
\bauthor{\bsnm{Lu}, \binits{H.-M.}},
\bauthor{\bsnm{Huang}, \binits{J.-H.}},
\bauthor{\bsnm{Hu}, \binits{Q.-N.}},
\bauthor{\bsnm{Liang}, \binits{Y.-Z.}}:
\batitle{Large-scale prediction of drug--target interactions using protein
  sequences and drug topological structures}.
\bjtitle{Analytica chimica acta}
\bvolume{752},
\bfpage{1}--\blpage{10}
(\byear{2012})
\end{barticle}
\endbibitem

%%% 59
\bibitem{friedman1997bias}
\begin{barticle}
\bauthor{\bsnm{Friedman}, \binits{J.H.}}:
\batitle{On bias, variance, 0/1—loss, and the curse-of-dimensionality}.
\bjtitle{Data mining and knowledge discovery}
\bvolume{1}(\bissue{1}),
\bfpage{55}--\blpage{77}
(\byear{1997})
\end{barticle}
\endbibitem

%%% 60
\bibitem{efron1983leisurely}
\begin{barticle}
\bauthor{\bsnm{Efron}, \binits{B.}},
\bauthor{\bsnm{Gong}, \binits{G.}}:
\batitle{A leisurely look at the bootstrap, the jackknife, and
  cross-validation}.
\bjtitle{The American Statistician}
\bvolume{37}(\bissue{1}),
\bfpage{36}--\blpage{48}
(\byear{1983})
\end{barticle}
\endbibitem

%%% 61
\bibitem{pedregosa2011scikit}
\begin{barticle}
\bauthor{\bsnm{Pedregosa}, \binits{F.}},
\bauthor{\bsnm{Varoquaux}, \binits{G.}},
\bauthor{\bsnm{Gramfort}, \binits{A.}},
\bauthor{\bsnm{Michel}, \binits{V.}},
\bauthor{\bsnm{Thirion}, \binits{B.}},
\bauthor{\bsnm{Grisel}, \binits{O.}},
\bauthor{\bsnm{Blondel}, \binits{M.}},
\bauthor{\bsnm{Prettenhofer}, \binits{P.}},
\bauthor{\bsnm{Weiss}, \binits{R.}},
\bauthor{\bsnm{Dubourg}, \binits{V.}}, \betal:
\batitle{Scikit-learn: Machine learning in python}.
\bjtitle{Journal of Machine Learning Research}
\bvolume{12}(\bissue{Oct}),
\bfpage{2825}--\blpage{2830}
(\byear{2011})
\end{barticle}
\endbibitem

%%% 62
\bibitem{ho1998random}
\begin{barticle}
\bauthor{\bsnm{Ho}, \binits{T.K.}}:
\batitle{The random subspace method for constructing decision forests}.
\bjtitle{IEEE transactions on pattern analysis and machine intelligence}
\bvolume{20}(\bissue{8}),
\bfpage{832}--\blpage{844}
(\byear{1998})
\end{barticle}
\endbibitem

%%% 63
\bibitem{cortes1995support}
\begin{barticle}
\bauthor{\bsnm{Cortes}, \binits{C.}},
\bauthor{\bsnm{Vapnik}, \binits{V.}}:
\batitle{Support-vector networks}.
\bjtitle{Machine learning}
\bvolume{20}(\bissue{3}),
\bfpage{273}--\blpage{297}
(\byear{1995})
\end{barticle}
\endbibitem

%%% 64
\bibitem{nanni2014set}
\begin{barticle}
\bauthor{\bsnm{Nanni}, \binits{L.}},
\bauthor{\bsnm{Lumini}, \binits{A.}},
\bauthor{\bsnm{Brahnam}, \binits{S.}}:
\batitle{A set of descriptors for identifying the protein--drug interaction in
  cellular networking}.
\bjtitle{Journal of theoretical biology}
\bvolume{359},
\bfpage{120}--\blpage{128}
(\byear{2014})
\end{barticle}
\endbibitem

\end{thebibliography}

\newcommand{\BMCxmlcomment}[1]{}

\BMCxmlcomment{

<refgrp>

<bibl id="B1">
  <title><p>Predicting new molecular targets for known drugs</p></title>
  <aug>
    <au><snm>Keiser</snm><fnm>MJ</fnm></au>
    <au><snm>Setola</snm><fnm>V</fnm></au>
    <au><snm>Irwin</snm><fnm>JJ</fnm></au>
    <au><snm>Laggner</snm><fnm>C</fnm></au>
    <au><snm>Abbas</snm><fnm>AI</fnm></au>
    <au><snm>Hufeisen</snm><fnm>SJ</fnm></au>
    <au><snm>Jensen</snm><fnm>NH</fnm></au>
    <au><snm>Kuijer</snm><fnm>MB</fnm></au>
    <au><snm>Matos</snm><fnm>RC</fnm></au>
    <au><snm>Tran</snm><fnm>TB</fnm></au>
    <au><cnm>others</cnm></au>
  </aug>
  <source>Nature</source>
  <publisher>Nature Publishing Group</publisher>
  <pubdate>2009</pubdate>
  <volume>462</volume>
  <issue>7270</issue>
  <fpage>175</fpage>
  <lpage>-181</lpage>
</bibl>

<bibl id="B2">
  <title><p>Prediction of polypharmacological profiles of drugs by the
  integration of chemical, side effect, and therapeutic space</p></title>
  <aug>
    <au><snm>Cheng</snm><fnm>F</fnm></au>
    <au><snm>Li</snm><fnm>W</fnm></au>
    <au><snm>Wu</snm><fnm>Z</fnm></au>
    <au><snm>Wang</snm><fnm>X</fnm></au>
    <au><snm>Zhang</snm><fnm>C</fnm></au>
    <au><snm>Li</snm><fnm>J</fnm></au>
    <au><snm>Liu</snm><fnm>G</fnm></au>
    <au><snm>Tang</snm><fnm>Y</fnm></au>
  </aug>
  <source>Journal of chemical information and modeling</source>
  <publisher>ACS Publications</publisher>
  <pubdate>2013</pubdate>
  <volume>53</volume>
  <issue>4</issue>
  <fpage>753</fpage>
  <lpage>-762</lpage>
</bibl>

<bibl id="B3">
  <title><p>SDTNBI: an integrated network and chemoinformatics tool for
  systematic prediction of drug--target interactions and drug
  repositioning</p></title>
  <aug>
    <au><snm>Wu</snm><fnm>Z</fnm></au>
    <au><snm>Cheng</snm><fnm>F</fnm></au>
    <au><snm>Li</snm><fnm>J</fnm></au>
    <au><snm>Li</snm><fnm>W</fnm></au>
    <au><snm>Liu</snm><fnm>G</fnm></au>
    <au><snm>Tang</snm><fnm>Y</fnm></au>
  </aug>
  <source>Briefings in bioinformatics</source>
  <publisher>Oxford University Press</publisher>
  <pubdate>2017</pubdate>
  <volume>18</volume>
  <issue>2</issue>
  <fpage>333</fpage>
  <lpage>-347</lpage>
</bibl>

<bibl id="B4">
  <title><p>Drug target identification using side-effect similarity</p></title>
  <aug>
    <au><snm>Campillos</snm><fnm>M</fnm></au>
    <au><snm>Kuhn</snm><fnm>M</fnm></au>
    <au><snm>Gavin</snm><fnm>AC</fnm></au>
    <au><snm>Jensen</snm><fnm>LJ</fnm></au>
    <au><snm>Bork</snm><fnm>P</fnm></au>
  </aug>
  <source>Science</source>
  <publisher>American Association for the Advancement of Science</publisher>
  <pubdate>2008</pubdate>
  <volume>321</volume>
  <issue>5886</issue>
  <fpage>263</fpage>
  <lpage>-266</lpage>
</bibl>

<bibl id="B5">
  <title><p>Multidimensional chemical genetic analysis of diversity-oriented
  synthesis-derived deacetylase inhibitors using cell-based assays</p></title>
  <aug>
    <au><snm>Haggarty</snm><fnm>SJ</fnm></au>
    <au><snm>Koeller</snm><fnm>KM</fnm></au>
    <au><snm>Wong</snm><fnm>JC</fnm></au>
    <au><snm>Butcher</snm><fnm>RA</fnm></au>
    <au><snm>Schreiber</snm><fnm>SL</fnm></au>
  </aug>
  <source>Chemistry \& biology</source>
  <publisher>Elsevier</publisher>
  <pubdate>2003</pubdate>
  <volume>10</volume>
  <issue>5</issue>
  <fpage>383</fpage>
  <lpage>-396</lpage>
</bibl>

<bibl id="B6">
  <title><p>Dissecting glucose signalling with diversity-oriented synthesis and
  small-molecule microarrays</p></title>
  <aug>
    <au><snm>Kuruvilla</snm><fnm>FG</fnm></au>
    <au><snm>Shamji</snm><fnm>AF</fnm></au>
    <au><snm>Sternson</snm><fnm>SM</fnm></au>
    <au><snm>Hergenrother</snm><fnm>PJ</fnm></au>
    <au><snm>Schreiber</snm><fnm>SL</fnm></au>
  </aug>
  <source>Nature</source>
  <publisher>Nature Publishing Group</publisher>
  <pubdate>2002</pubdate>
  <volume>416</volume>
  <issue>6881</issue>
  <fpage>653</fpage>
  <lpage>-657</lpage>
</bibl>

<bibl id="B7">
  <title><p>The role of ligand efficiency metrics in drug discovery</p></title>
  <aug>
    <au><snm>Hopkins</snm><fnm>AL</fnm></au>
    <au><snm>Keser{\"u}</snm><fnm>GM</fnm></au>
    <au><snm>Leeson</snm><fnm>PD</fnm></au>
    <au><snm>Rees</snm><fnm>DC</fnm></au>
    <au><snm>Reynolds</snm><fnm>CH</fnm></au>
  </aug>
  <source>Nature Reviews Drug Discovery</source>
  <publisher>Nature Research</publisher>
  <pubdate>2014</pubdate>
  <volume>13</volume>
  <issue>2</issue>
  <fpage>105</fpage>
  <lpage>-121</lpage>
</bibl>

<bibl id="B8">
  <title><p>Relating protein pharmacology by ligand chemistry</p></title>
  <aug>
    <au><snm>Keiser</snm><fnm>MJ</fnm></au>
    <au><snm>Roth</snm><fnm>BL</fnm></au>
    <au><snm>Armbruster</snm><fnm>BN</fnm></au>
    <au><snm>Ernsberger</snm><fnm>P</fnm></au>
    <au><snm>Irwin</snm><fnm>JJ</fnm></au>
    <au><snm>Shoichet</snm><fnm>BK</fnm></au>
  </aug>
  <source>Nature biotechnology</source>
  <publisher>Nature Publishing Group</publisher>
  <pubdate>2007</pubdate>
  <volume>25</volume>
  <issue>2</issue>
  <fpage>197</fpage>
  <lpage>-206</lpage>
</bibl>

<bibl id="B9">
  <title><p>Drug repositioning by structure-based virtual screening</p></title>
  <aug>
    <au><snm>Ma</snm><fnm>DL</fnm></au>
    <au><snm>Chan</snm><fnm>DSH</fnm></au>
    <au><snm>Leung</snm><fnm>CH</fnm></au>
  </aug>
  <source>Chemical Society Reviews</source>
  <publisher>Royal Society of Chemistry</publisher>
  <pubdate>2013</pubdate>
  <volume>42</volume>
  <issue>5</issue>
  <fpage>2130</fpage>
  <lpage>-2141</lpage>
</bibl>

<bibl id="B10">
  <title><p>Molecular determinants of drug--receptor binding
  kinetics</p></title>
  <aug>
    <au><snm>Pan</snm><fnm>AC</fnm></au>
    <au><snm>Borhani</snm><fnm>DW</fnm></au>
    <au><snm>Dror</snm><fnm>RO</fnm></au>
    <au><snm>Shaw</snm><fnm>DE</fnm></au>
  </aug>
  <source>Drug discovery today</source>
  <publisher>Elsevier</publisher>
  <pubdate>2013</pubdate>
  <volume>18</volume>
  <issue>13</issue>
  <fpage>667</fpage>
  <lpage>-673</lpage>
</bibl>

<bibl id="B11">
  <title><p>A drug target slim: using gene ontology and gene ontology
  annotations to navigate protein-ligand target space in ChEMBL</p></title>
  <aug>
    <au><snm>Mutowo</snm><fnm>P</fnm></au>
    <au><snm>Bento</snm><fnm>AP</fnm></au>
    <au><snm>Dedman</snm><fnm>N</fnm></au>
    <au><snm>Gaulton</snm><fnm>A</fnm></au>
    <au><snm>Hersey</snm><fnm>A</fnm></au>
    <au><snm>Lomax</snm><fnm>J</fnm></au>
    <au><snm>Overington</snm><fnm>JP</fnm></au>
  </aug>
  <source>Journal of biomedical semantics</source>
  <publisher>BioMed Central</publisher>
  <pubdate>2016</pubdate>
  <volume>7</volume>
  <issue>1</issue>
  <fpage>59</fpage>
</bibl>

<bibl id="B12">
  <title><p>Computational polypharmacology with text mining and
  ontologies</p></title>
  <aug>
    <au><snm>Plake</snm><fnm>C</fnm></au>
    <au><snm>Schroeder</snm><fnm>M</fnm></au>
  </aug>
  <source>Current pharmaceutical biotechnology</source>
  <publisher>Bentham Science Publishers</publisher>
  <pubdate>2011</pubdate>
  <volume>12</volume>
  <issue>3</issue>
  <fpage>449</fpage>
  <lpage>-457</lpage>
</bibl>

<bibl id="B13">
  <title><p>A probabilistic model for mining implicit ‘chemical
  compound--gene’relations from literature</p></title>
  <aug>
    <au><snm>Zhu</snm><fnm>S</fnm></au>
    <au><snm>Okuno</snm><fnm>Y</fnm></au>
    <au><snm>Tsujimoto</snm><fnm>G</fnm></au>
    <au><snm>Mamitsuka</snm><fnm>H</fnm></au>
  </aug>
  <source>Bioinformatics</source>
  <publisher>Oxford Univ Press</publisher>
  <pubdate>2005</pubdate>
  <volume>21</volume>
  <issue>suppl 2</issue>
  <fpage>ii245</fpage>
  <lpage>-ii251</lpage>
</bibl>

<bibl id="B14">
  <title><p>AutoDock4 and AutoDockTools4: Automated docking with selective
  receptor flexibility</p></title>
  <aug>
    <au><snm>Morris</snm><fnm>GM</fnm></au>
    <au><snm>Huey</snm><fnm>R</fnm></au>
    <au><snm>Lindstrom</snm><fnm>W</fnm></au>
    <au><snm>Sanner</snm><fnm>MF</fnm></au>
    <au><snm>Belew</snm><fnm>RK</fnm></au>
    <au><snm>Goodsell</snm><fnm>DS</fnm></au>
    <au><snm>Olson</snm><fnm>AJ</fnm></au>
  </aug>
  <source>Journal of computational chemistry</source>
  <publisher>Wiley Online Library</publisher>
  <pubdate>2009</pubdate>
  <volume>30</volume>
  <issue>16</issue>
  <fpage>2785</fpage>
  <lpage>-2791</lpage>
</bibl>

<bibl id="B15">
  <title><p>Drug--target interaction prediction via chemogenomic space:
  learning-based methods</p></title>
  <aug>
    <au><snm>Mousavian</snm><fnm>Z</fnm></au>
    <au><snm>Masoudi Nejad</snm><fnm>A</fnm></au>
  </aug>
  <source>Expert opinion on drug metabolism \& toxicology</source>
  <publisher>Taylor \& Francis</publisher>
  <pubdate>2014</pubdate>
  <volume>10</volume>
  <issue>9</issue>
  <fpage>1273</fpage>
  <lpage>-1287</lpage>
</bibl>

<bibl id="B16">
  <title><p>Prediction of drug--target interaction networks from the
  integration of chemical and genomic spaces</p></title>
  <aug>
    <au><snm>Yamanishi</snm><fnm>Y</fnm></au>
    <au><snm>Araki</snm><fnm>M</fnm></au>
    <au><snm>Gutteridge</snm><fnm>A</fnm></au>
    <au><snm>Honda</snm><fnm>W</fnm></au>
    <au><snm>Kanehisa</snm><fnm>M</fnm></au>
  </aug>
  <source>Bioinformatics</source>
  <publisher>Oxford Univ Press</publisher>
  <pubdate>2008</pubdate>
  <volume>24</volume>
  <issue>13</issue>
  <fpage>i232</fpage>
  <lpage>-i240</lpage>
</bibl>

<bibl id="B17">
  <title><p>Supervised prediction of drug--target interactions using bipartite
  local models</p></title>
  <aug>
    <au><snm>Bleakley</snm><fnm>K</fnm></au>
    <au><snm>Yamanishi</snm><fnm>Y</fnm></au>
  </aug>
  <source>Bioinformatics</source>
  <publisher>Oxford Univ Press</publisher>
  <pubdate>2009</pubdate>
  <volume>25</volume>
  <issue>18</issue>
  <fpage>2397</fpage>
  <lpage>-2403</lpage>
</bibl>

<bibl id="B18">
  <title><p>Drug target predictions based on heterogeneous graph
  inference</p></title>
  <aug>
    <au><snm>Wang</snm><fnm>W</fnm></au>
    <au><snm>Yang</snm><fnm>S</fnm></au>
    <au><snm>Li</snm><fnm>JING</fnm></au>
  </aug>
  <source>Pacific Symposium on Biocomputing. Pacific Symposium on
  Biocomputing</source>
  <pubdate>2013</pubdate>
  <fpage>53</fpage>
</bibl>

<bibl id="B19">
  <title><p>Drug--target interaction prediction by random walk on the
  heterogeneous network</p></title>
  <aug>
    <au><snm>Chen</snm><fnm>X</fnm></au>
    <au><snm>Liu</snm><fnm>MX</fnm></au>
    <au><snm>Yan</snm><fnm>GY</fnm></au>
  </aug>
  <source>Molecular BioSystems</source>
  <publisher>Royal Society of Chemistry</publisher>
  <pubdate>2012</pubdate>
  <volume>8</volume>
  <issue>7</issue>
  <fpage>1970</fpage>
  <lpage>-1978</lpage>
</bibl>

<bibl id="B20">
  <title><p>Drug--target interaction prediction through domain-tuned
  network-based inference</p></title>
  <aug>
    <au><snm>Alaimo</snm><fnm>S</fnm></au>
    <au><snm>Pulvirenti</snm><fnm>A</fnm></au>
    <au><snm>Giugno</snm><fnm>R</fnm></au>
    <au><snm>Ferro</snm><fnm>A</fnm></au>
  </aug>
  <source>Bioinformatics</source>
  <publisher>Oxford Univ Press</publisher>
  <pubdate>2013</pubdate>
  <volume>29</volume>
  <issue>16</issue>
  <fpage>2004</fpage>
  <lpage>-2008</lpage>
</bibl>

<bibl id="B21">
  <title><p>Prediction of drug-target interactions and drug repositioning via
  network-based inference</p></title>
  <aug>
    <au><snm>Cheng</snm><fnm>F</fnm></au>
    <au><snm>Liu</snm><fnm>C</fnm></au>
    <au><snm>Jiang</snm><fnm>J</fnm></au>
    <au><snm>Lu</snm><fnm>W</fnm></au>
    <au><snm>Li</snm><fnm>W</fnm></au>
    <au><snm>Liu</snm><fnm>G</fnm></au>
    <au><snm>Zhou</snm><fnm>W</fnm></au>
    <au><snm>Huang</snm><fnm>J</fnm></au>
    <au><snm>Tang</snm><fnm>Y</fnm></au>
  </aug>
  <source>PLoS Comput Biol</source>
  <publisher>Public Library of Science</publisher>
  <pubdate>2012</pubdate>
  <volume>8</volume>
  <issue>5</issue>
  <fpage>e1002503</fpage>
</bibl>

<bibl id="B22">
  <title><p>Drug--target interaction prediction from PSSM based evolutionary
  information</p></title>
  <aug>
    <au><snm>Mousavian</snm><fnm>Z</fnm></au>
    <au><snm>Khakabimamaghani</snm><fnm>S</fnm></au>
    <au><snm>Kavousi</snm><fnm>K</fnm></au>
    <au><snm>Masoudi Nejad</snm><fnm>A</fnm></au>
  </aug>
  <source>Journal of pharmacological and toxicological methods</source>
  <publisher>Elsevier</publisher>
  <pubdate>2016</pubdate>
  <volume>78</volume>
  <fpage>42</fpage>
  <lpage>-51</lpage>
</bibl>

<bibl id="B23">
  <title><p>SELF-BLM: Prediction of drug-target interactions via self-training
  SVM</p></title>
  <aug>
    <au><snm>Keum</snm><fnm>J</fnm></au>
    <au><snm>Nam</snm><fnm>H</fnm></au>
  </aug>
  <source>PloS one</source>
  <publisher>Public Library of Science</publisher>
  <pubdate>2017</pubdate>
  <volume>12</volume>
  <issue>2</issue>
  <fpage>e0171839</fpage>
</bibl>

<bibl id="B24">
  <title><p>Large-scale prediction of drug-target interactions from deep
  representations</p></title>
  <aug>
    <au><snm>Chan</snm><fnm>KC</fnm></au>
    <au><snm>You</snm><fnm>ZH</fnm></au>
    <au><cnm>others</cnm></au>
  </aug>
  <source>Neural Networks (IJCNN), 2016 International Joint Conference
  on</source>
  <pubdate>2016</pubdate>
  <fpage>1236</fpage>
  <lpage>-1243</lpage>
</bibl>

<bibl id="B25">
  <title><p>iCDI-PseFpt: identify the channel--drug interaction in cellular
  networking with PseAAC and molecular fingerprints</p></title>
  <aug>
    <au><snm>Xiao</snm><fnm>X</fnm></au>
    <au><snm>Min</snm><fnm>JL</fnm></au>
    <au><snm>Wang</snm><fnm>P</fnm></au>
    <au><snm>Chou</snm><fnm>KC</fnm></au>
  </aug>
  <source>Journal of theoretical biology</source>
  <publisher>Elsevier</publisher>
  <pubdate>2013</pubdate>
  <volume>337</volume>
  <fpage>71</fpage>
  <lpage>-79</lpage>
</bibl>

<bibl id="B26">
  <title><p>Predicting drug-target interaction networks based on functional
  groups and biological features</p></title>
  <aug>
    <au><snm>He</snm><fnm>Z</fnm></au>
    <au><snm>Zhang</snm><fnm>J</fnm></au>
    <au><snm>Shi</snm><fnm>XH</fnm></au>
    <au><snm>Hu</snm><fnm>LL</fnm></au>
    <au><snm>Kong</snm><fnm>X</fnm></au>
    <au><snm>Cai</snm><fnm>YD</fnm></au>
    <au><snm>Chou</snm><fnm>KC</fnm></au>
  </aug>
  <source>PloS one</source>
  <publisher>Public Library of Science</publisher>
  <pubdate>2010</pubdate>
  <volume>5</volume>
  <issue>3</issue>
  <fpage>e9603</fpage>
</bibl>

<bibl id="B27">
  <title><p>Drug-target interaction prediction from chemical, genomic and
  pharmacological data in an integrated framework</p></title>
  <aug>
    <au><snm>Yamanishi</snm><fnm>Y</fnm></au>
    <au><snm>Kotera</snm><fnm>M</fnm></au>
    <au><snm>Kanehisa</snm><fnm>M</fnm></au>
    <au><snm>Goto</snm><fnm>S</fnm></au>
  </aug>
  <source>Bioinformatics</source>
  <publisher>Oxford Univ Press</publisher>
  <pubdate>2010</pubdate>
  <volume>26</volume>
  <issue>12</issue>
  <fpage>i246</fpage>
  <lpage>-i254</lpage>
</bibl>

<bibl id="B28">
  <title><p>Improved prediction of drug-target interactions using regularized
  least squares integrating with kernel fusion technique</p></title>
  <aug>
    <au><snm>Hao</snm><fnm>M</fnm></au>
    <au><snm>Wang</snm><fnm>Y</fnm></au>
    <au><snm>Bryant</snm><fnm>SH</fnm></au>
  </aug>
  <source>Analytica chimica acta</source>
  <publisher>Elsevier</publisher>
  <pubdate>2016</pubdate>
  <volume>909</volume>
  <fpage>41</fpage>
  <lpage>-50</lpage>
</bibl>

<bibl id="B29">
  <title><p>Predicting drug--target interactions from chemical and genomic
  kernels using Bayesian matrix factorization</p></title>
  <aug>
    <au><snm>G{\"o}nen</snm><fnm>M</fnm></au>
  </aug>
  <source>Bioinformatics</source>
  <publisher>Oxford Univ Press</publisher>
  <pubdate>2012</pubdate>
  <volume>28</volume>
  <issue>18</issue>
  <fpage>2304</fpage>
  <lpage>-2310</lpage>
</bibl>

<bibl id="B30">
  <title><p>A semi-supervised method for drug-target interaction prediction
  with consistency in networks</p></title>
  <aug>
    <au><snm>Chen</snm><fnm>H</fnm></au>
    <au><snm>Zhang</snm><fnm>Z</fnm></au>
  </aug>
  <source>PloS one</source>
  <publisher>Public Library of Science</publisher>
  <pubdate>2013</pubdate>
  <volume>8</volume>
  <issue>5</issue>
  <fpage>e62975</fpage>
</bibl>

<bibl id="B31">
  <title><p>DASPfind: new efficient method to predict drug--target
  interactions</p></title>
  <aug>
    <au><snm>Ba Alawi</snm><fnm>W</fnm></au>
    <au><snm>Soufan</snm><fnm>O</fnm></au>
    <au><snm>Essack</snm><fnm>M</fnm></au>
    <au><snm>Kalnis</snm><fnm>P</fnm></au>
    <au><snm>Bajic</snm><fnm>VB</fnm></au>
  </aug>
  <source>Journal of cheminformatics</source>
  <publisher>Springer International Publishing</publisher>
  <pubdate>2016</pubdate>
  <volume>8</volume>
  <issue>1</issue>
  <fpage>15</fpage>
</bibl>

<bibl id="B32">
  <title><p>Sixty-five years of the long march in protein secondary structure
  prediction: the final stretch?</p></title>
  <aug>
    <au><snm>Yang</snm><fnm>Y</fnm></au>
    <au><snm>Gao</snm><fnm>J</fnm></au>
    <au><snm>Wang</snm><fnm>J</fnm></au>
    <au><snm>Heffernan</snm><fnm>R</fnm></au>
    <au><snm>Hanson</snm><fnm>J</fnm></au>
    <au><snm>Paliwal</snm><fnm>K</fnm></au>
    <au><snm>Zhou</snm><fnm>Y</fnm></au>
  </aug>
  <source>Briefings in Bioinformatics</source>
  <publisher>Oxford Univ Press</publisher>
  <pubdate>2016</pubdate>
  <fpage>bbw129</fpage>
</bibl>

<bibl id="B33">
  <title><p>SPIDER2: A Package to Predict Secondary Structure, Accessible
  Surface Area, and Main-Chain Torsional Angles by Deep Neural
  Networks</p></title>
  <aug>
    <au><snm>Yang</snm><fnm>Y</fnm></au>
    <au><snm>Heffernan</snm><fnm>R</fnm></au>
    <au><snm>Paliwal</snm><fnm>K</fnm></au>
    <au><snm>Lyons</snm><fnm>J</fnm></au>
    <au><snm>Dehzangi</snm><fnm>A</fnm></au>
    <au><snm>Sharma</snm><fnm>A</fnm></au>
    <au><snm>Wang</snm><fnm>J</fnm></au>
    <au><snm>Sattar</snm><fnm>A</fnm></au>
    <au><snm>Zhou</snm><fnm>Y</fnm></au>
  </aug>
  <source>Prediction of Protein Secondary Structure</source>
  <publisher>Springer</publisher>
  <pubdate>2017</pubdate>
  <fpage>55</fpage>
  <lpage>-63</lpage>
</bibl>

<bibl id="B34">
  <title><p>SucStruct: Prediction of succinylated lysine residues by using
  structural properties of amino acids</p></title>
  <aug>
    <au><snm>L{\'o}pez</snm><fnm>Y</fnm></au>
    <au><snm>Dehzangi</snm><fnm>A</fnm></au>
    <au><snm>Lal</snm><fnm>SP</fnm></au>
    <au><snm>Taherzadeh</snm><fnm>G</fnm></au>
    <au><snm>Michaelson</snm><fnm>J</fnm></au>
    <au><snm>Sattar</snm><fnm>A</fnm></au>
    <au><snm>Tsunoda</snm><fnm>T</fnm></au>
    <au><snm>Sharma</snm><fnm>A</fnm></au>
  </aug>
  <source>Analytical Biochemistry</source>
  <publisher>Elsevier</publisher>
  <pubdate>2017</pubdate>
</bibl>

<bibl id="B35">
  <title><p>Some remarks on protein attribute prediction and pseudo amino acid
  composition</p></title>
  <aug>
    <au><snm>Chou</snm><fnm>KC</fnm></au>
  </aug>
  <source>Journal of theoretical biology</source>
  <publisher>Elsevier</publisher>
  <pubdate>2011</pubdate>
  <volume>273</volume>
  <issue>1</issue>
  <fpage>236</fpage>
  <lpage>-247</lpage>
</bibl>

<bibl id="B36">
  <title><p>DrugBank 3.0: a comprehensive resource for ‘omics’ research on
  drugs</p></title>
  <aug>
    <au><snm>Knox</snm><fnm>C</fnm></au>
    <au><snm>Law</snm><fnm>V</fnm></au>
    <au><snm>Jewison</snm><fnm>T</fnm></au>
    <au><snm>Liu</snm><fnm>P</fnm></au>
    <au><snm>Ly</snm><fnm>S</fnm></au>
    <au><snm>Frolkis</snm><fnm>A</fnm></au>
    <au><snm>Pon</snm><fnm>A</fnm></au>
    <au><snm>Banco</snm><fnm>K</fnm></au>
    <au><snm>Mak</snm><fnm>C</fnm></au>
    <au><snm>Neveu</snm><fnm>V</fnm></au>
    <au><cnm>others</cnm></au>
  </aug>
  <source>Nucleic acids research</source>
  <publisher>Oxford Univ Press</publisher>
  <pubdate>2011</pubdate>
  <volume>39</volume>
  <issue>suppl 1</issue>
  <fpage>D1035</fpage>
  <lpage>-D1041</lpage>
</bibl>

<bibl id="B37">
  <title><p>KEGG: kyoto encyclopedia of genes and genomes</p></title>
  <aug>
    <au><snm>Kanehisa</snm><fnm>M</fnm></au>
    <au><snm>Goto</snm><fnm>S</fnm></au>
  </aug>
  <source>Nucleic acids research</source>
  <publisher>Oxford Univ Press</publisher>
  <pubdate>2000</pubdate>
  <volume>28</volume>
  <issue>1</issue>
  <fpage>27</fpage>
  <lpage>-30</lpage>
</bibl>

<bibl id="B38">
  <title><p>Gapped BLAST and PSI-BLAST: a new generation of protein database
  search programs</p></title>
  <aug>
    <au><snm>Altschul</snm><fnm>SF</fnm></au>
    <au><snm>Madden</snm><fnm>TL</fnm></au>
    <au><snm>Sch{\"a}ffer</snm><fnm>AA</fnm></au>
    <au><snm>Zhang</snm><fnm>J</fnm></au>
    <au><snm>Zhang</snm><fnm>Z</fnm></au>
    <au><snm>Miller</snm><fnm>W</fnm></au>
    <au><snm>Lipman</snm><fnm>DJ</fnm></au>
  </aug>
  <source>Nucleic acids research</source>
  <publisher>Oxford Univ Press</publisher>
  <pubdate>1997</pubdate>
  <volume>25</volume>
  <issue>17</issue>
  <fpage>3389</fpage>
  <lpage>-3402</lpage>
</bibl>

<bibl id="B39">
  <title><p>DrugBank: a knowledgebase for drugs, drug actions and drug
  targets</p></title>
  <aug>
    <au><snm>Wishart</snm><fnm>DS</fnm></au>
    <au><snm>Knox</snm><fnm>C</fnm></au>
    <au><snm>Guo</snm><fnm>AC</fnm></au>
    <au><snm>Cheng</snm><fnm>D</fnm></au>
    <au><snm>Shrivastava</snm><fnm>S</fnm></au>
    <au><snm>Tzur</snm><fnm>D</fnm></au>
    <au><snm>Gautam</snm><fnm>B</fnm></au>
    <au><snm>Hassanali</snm><fnm>M</fnm></au>
  </aug>
  <source>Nucleic acids research</source>
  <publisher>Oxford Univ Press</publisher>
  <pubdate>2008</pubdate>
  <volume>36</volume>
  <issue>suppl 1</issue>
  <fpage>D901</fpage>
  <lpage>-D906</lpage>
</bibl>

<bibl id="B40">
  <title><p>KEGG for linking genomes to life and the environment</p></title>
  <aug>
    <au><snm>Kanehisa</snm><fnm>M</fnm></au>
    <au><snm>Araki</snm><fnm>M</fnm></au>
    <au><snm>Goto</snm><fnm>S</fnm></au>
    <au><snm>Hattori</snm><fnm>M</fnm></au>
    <au><snm>Hirakawa</snm><fnm>M</fnm></au>
    <au><snm>Itoh</snm><fnm>M</fnm></au>
    <au><snm>Katayama</snm><fnm>T</fnm></au>
    <au><snm>Kawashima</snm><fnm>S</fnm></au>
    <au><snm>Okuda</snm><fnm>S</fnm></au>
    <au><snm>Tokimatsu</snm><fnm>T</fnm></au>
    <au><cnm>others</cnm></au>
  </aug>
  <source>Nucleic acids research</source>
  <publisher>Oxford Univ Press</publisher>
  <pubdate>2008</pubdate>
  <volume>36</volume>
  <issue>suppl 1</issue>
  <fpage>D480</fpage>
  <lpage>-D484</lpage>
</bibl>

<bibl id="B41">
  <title><p>BRENDA, the enzyme database: updates and major new
  developments</p></title>
  <aug>
    <au><snm>Schomburg</snm><fnm>I</fnm></au>
    <au><snm>Chang</snm><fnm>A</fnm></au>
    <au><snm>Ebeling</snm><fnm>C</fnm></au>
    <au><snm>Gremse</snm><fnm>M</fnm></au>
    <au><snm>Heldt</snm><fnm>C</fnm></au>
    <au><snm>Huhn</snm><fnm>G</fnm></au>
    <au><snm>Schomburg</snm><fnm>D</fnm></au>
  </aug>
  <source>Nucleic acids research</source>
  <publisher>Oxford Univ Press</publisher>
  <pubdate>2004</pubdate>
  <volume>32</volume>
  <issue>suppl 1</issue>
  <fpage>D431</fpage>
  <lpage>-D433</lpage>
</bibl>

<bibl id="B42">
  <title><p>SuperTarget and Matador: resources for exploring drug-target
  relationships</p></title>
  <aug>
    <au><snm>G{\"u}nther</snm><fnm>S</fnm></au>
    <au><snm>Kuhn</snm><fnm>M</fnm></au>
    <au><snm>Dunkel</snm><fnm>M</fnm></au>
    <au><snm>Campillos</snm><fnm>M</fnm></au>
    <au><snm>Senger</snm><fnm>C</fnm></au>
    <au><snm>Petsalaki</snm><fnm>E</fnm></au>
    <au><snm>Ahmed</snm><fnm>J</fnm></au>
    <au><snm>Urdiales</snm><fnm>EG</fnm></au>
    <au><snm>Gewiess</snm><fnm>A</fnm></au>
    <au><snm>Jensen</snm><fnm>LJ</fnm></au>
    <au><cnm>others</cnm></au>
  </aug>
  <source>Nucleic acids research</source>
  <publisher>Oxford Univ Press</publisher>
  <pubdate>2008</pubdate>
  <volume>36</volume>
  <issue>suppl 1</issue>
  <fpage>D919</fpage>
  <lpage>-D922</lpage>
</bibl>

<bibl id="B43">
  <title><p>Handbook of molecular descriptors</p></title>
  <aug>
    <au><snm>Todeschini</snm><fnm>R</fnm></au>
    <au><snm>Consonni</snm><fnm>V</fnm></au>
  </aug>
  <publisher>John Wiley \& Sons</publisher>
  <pubdate>2008</pubdate>
  <volume>11</volume>
</bibl>

<bibl id="B44">
  <title><p>Scalable prediction of compound-protein interactions using minwise
  hashing</p></title>
  <aug>
    <au><snm>Tabei</snm><fnm>Y</fnm></au>
    <au><snm>Yamanishi</snm><fnm>Y</fnm></au>
  </aug>
  <source>BMC systems biology</source>
  <publisher>BioMed Central</publisher>
  <pubdate>2013</pubdate>
  <volume>7</volume>
  <issue>6</issue>
  <fpage>S3</fpage>
</bibl>

<bibl id="B45">
  <title><p>Identification of chemogenomic features from drug--target
  interaction networks using interpretable classifiers</p></title>
  <aug>
    <au><snm>Tabei</snm><fnm>Y</fnm></au>
    <au><snm>Pauwels</snm><fnm>E</fnm></au>
    <au><snm>Stoven</snm><fnm>V</fnm></au>
    <au><snm>Takemoto</snm><fnm>K</fnm></au>
    <au><snm>Yamanishi</snm><fnm>Y</fnm></au>
  </aug>
  <source>Bioinformatics</source>
  <publisher>Oxford Univ Press</publisher>
  <pubdate>2012</pubdate>
  <volume>28</volume>
  <issue>18</issue>
  <fpage>i487</fpage>
  <lpage>-i494</lpage>
</bibl>

<bibl id="B46">
  <title><p>PubChem as a source of polypharmacology</p></title>
  <aug>
    <au><snm>Chen</snm><fnm>B</fnm></au>
    <au><snm>Wild</snm><fnm>D</fnm></au>
    <au><snm>Guha</snm><fnm>R</fnm></au>
  </aug>
  <source>Journal of chemical information and modeling</source>
  <publisher>ACS Publications</publisher>
  <pubdate>2009</pubdate>
  <volume>49</volume>
  <issue>9</issue>
  <fpage>2044</fpage>
  <lpage>-2055</lpage>
</bibl>

<bibl id="B47">
  <title><p>Chemical informatics functionality in R</p></title>
  <aug>
    <au><snm>Guha</snm><fnm>R</fnm></au>
    <au><cnm>others</cnm></au>
  </aug>
  <source>Journal of Statistical Software</source>
  <pubdate>2007</pubdate>
  <volume>18</volume>
  <issue>5</issue>
  <fpage>1</fpage>
  <lpage>-16</lpage>
</bibl>

<bibl id="B48">
  <title><p>Predict Gram-positive and gram-negative subcellular localization
  via incorporating evolutionary information and physicochemical features into
  Chou's general PseAAC</p></title>
  <aug>
    <au><snm>Sharma</snm><fnm>R</fnm></au>
    <au><snm>Dehzangi</snm><fnm>A</fnm></au>
    <au><snm>Lyons</snm><fnm>J</fnm></au>
    <au><snm>Paliwal</snm><fnm>K</fnm></au>
    <au><snm>Tsunoda</snm><fnm>T</fnm></au>
    <au><snm>Sharma</snm><fnm>A</fnm></au>
  </aug>
  <source>IEEE Transactions on NanoBioscience</source>
  <publisher>IEEE</publisher>
  <pubdate>2015</pubdate>
  <volume>14</volume>
  <issue>8</issue>
  <fpage>915</fpage>
  <lpage>-926</lpage>
</bibl>

<bibl id="B49">
  <title><p>A tri-gram based feature extraction technique using linear
  probabilities of position specific scoring matrix for protein fold
  recognition</p></title>
  <aug>
    <au><snm>Paliwal</snm><fnm>KK</fnm></au>
    <au><snm>Sharma</snm><fnm>A</fnm></au>
    <au><snm>Lyons</snm><fnm>J</fnm></au>
    <au><snm>Dehzangi</snm><fnm>A</fnm></au>
  </aug>
  <source>IEEE Transactions on Nanobioscience</source>
  <publisher>IEEE</publisher>
  <pubdate>2014</pubdate>
  <volume>13</volume>
  <issue>1</issue>
  <fpage>44</fpage>
  <lpage>-50</lpage>
</bibl>

<bibl id="B50">
  <title><p>SMOTE: synthetic minority over-sampling technique</p></title>
  <aug>
    <au><snm>Chawla</snm><fnm>NV</fnm></au>
    <au><snm>Bowyer</snm><fnm>KW</fnm></au>
    <au><snm>Hall</snm><fnm>LO</fnm></au>
    <au><snm>Kegelmeyer</snm><fnm>WP</fnm></au>
  </aug>
  <source>Journal of artificial intelligence research</source>
  <pubdate>2002</pubdate>
  <volume>16</volume>
  <fpage>321</fpage>
  <lpage>-357</lpage>
</bibl>

<bibl id="B51">
  <title><p>Simple sequence-based kernels do not predict protein--protein
  interactions</p></title>
  <aug>
    <au><snm>Yu</snm><fnm>J</fnm></au>
    <au><snm>Guo</snm><fnm>M</fnm></au>
    <au><snm>Needham</snm><fnm>CJ</fnm></au>
    <au><snm>Huang</snm><fnm>Y</fnm></au>
    <au><snm>Cai</snm><fnm>L</fnm></au>
    <au><snm>Westhead</snm><fnm>DR</fnm></au>
  </aug>
  <source>Bioinformatics</source>
  <publisher>Oxford Univ Press</publisher>
  <pubdate>2010</pubdate>
  <volume>26</volume>
  <issue>20</issue>
  <fpage>2610</fpage>
  <lpage>-2614</lpage>
</bibl>

<bibl id="B52">
  <title><p>Improving identification of difficult small classes by balancing
  class distribution</p></title>
  <aug>
    <au><snm>Laurikkala</snm><fnm>J</fnm></au>
  </aug>
  <source>Conference on Artificial Intelligence in Medicine in Europe</source>
  <pubdate>2001</pubdate>
  <fpage>63</fpage>
  <lpage>-66</lpage>
</bibl>

<bibl id="B53">
  <title><p>Cluster-based under-sampling approaches for imbalanced data
  distributions</p></title>
  <aug>
    <au><snm>Yen</snm><fnm>SJ</fnm></au>
    <au><snm>Lee</snm><fnm>YS</fnm></au>
  </aug>
  <source>Expert Systems with Applications</source>
  <publisher>Elsevier</publisher>
  <pubdate>2009</pubdate>
  <volume>36</volume>
  <issue>3</issue>
  <fpage>5718</fpage>
  <lpage>-5727</lpage>
</bibl>

<bibl id="B54">
  <title><p>Cluster based under-sampling for unbalanced cardiovascular
  data</p></title>
  <aug>
    <au><snm>Rahman</snm><fnm>MM</fnm></au>
    <au><snm>Davis</snm><fnm>D</fnm></au>
  </aug>
  <source>Proceedings of the World Congress on Engineering</source>
  <pubdate>2013</pubdate>
  <volume>3</volume>
  <fpage>3</fpage>
  <lpage>-5</lpage>
</bibl>

<bibl id="B55">
  <title><p>A desicion-theoretic generalization of on-line learning and an
  application to boosting</p></title>
  <aug>
    <au><snm>Freund</snm><fnm>Y</fnm></au>
    <au><snm>Schapire</snm><fnm>RE</fnm></au>
  </aug>
  <source>European conference on computational learning theory</source>
  <pubdate>1995</pubdate>
  <fpage>23</fpage>
  <lpage>-37</lpage>
</bibl>

<bibl id="B56">
  <title><p>Foundations of machine learning</p></title>
  <aug>
    <au><snm>Mohri</snm><fnm>M</fnm></au>
    <au><snm>Rostamizadeh</snm><fnm>A</fnm></au>
    <au><snm>Talwalkar</snm><fnm>A</fnm></au>
  </aug>
  <publisher>MIT press</publisher>
  <pubdate>2012</pubdate>
</bibl>

<bibl id="B57">
  <title><p>Evaluation: from precision, recall and F-measure to ROC,
  informedness, markedness and correlation</p></title>
  <aug>
    <au><snm>Powers</snm><fnm>DM</fnm></au>
  </aug>
  <publisher>Bioinfo Publications</publisher>
  <pubdate>2011</pubdate>
</bibl>

<bibl id="B58">
  <title><p>Large-scale prediction of drug--target interactions using protein
  sequences and drug topological structures</p></title>
  <aug>
    <au><snm>Cao</snm><fnm>DS</fnm></au>
    <au><snm>Liu</snm><fnm>S</fnm></au>
    <au><snm>Xu</snm><fnm>QS</fnm></au>
    <au><snm>Lu</snm><fnm>HM</fnm></au>
    <au><snm>Huang</snm><fnm>JH</fnm></au>
    <au><snm>Hu</snm><fnm>QN</fnm></au>
    <au><snm>Liang</snm><fnm>YZ</fnm></au>
  </aug>
  <source>Analytica chimica acta</source>
  <publisher>Elsevier</publisher>
  <pubdate>2012</pubdate>
  <volume>752</volume>
  <fpage>1</fpage>
  <lpage>-10</lpage>
</bibl>

<bibl id="B59">
  <title><p>On bias, variance, 0/1—loss, and the
  curse-of-dimensionality</p></title>
  <aug>
    <au><snm>Friedman</snm><fnm>JH</fnm></au>
  </aug>
  <source>Data mining and knowledge discovery</source>
  <publisher>Springer</publisher>
  <pubdate>1997</pubdate>
  <volume>1</volume>
  <issue>1</issue>
  <fpage>55</fpage>
  <lpage>-77</lpage>
</bibl>

<bibl id="B60">
  <title><p>A leisurely look at the bootstrap, the jackknife, and
  cross-validation</p></title>
  <aug>
    <au><snm>Efron</snm><fnm>B</fnm></au>
    <au><snm>Gong</snm><fnm>G</fnm></au>
  </aug>
  <source>The American Statistician</source>
  <publisher>Taylor \& Francis</publisher>
  <pubdate>1983</pubdate>
  <volume>37</volume>
  <issue>1</issue>
  <fpage>36</fpage>
  <lpage>-48</lpage>
</bibl>

<bibl id="B61">
  <title><p>Scikit-learn: Machine learning in Python</p></title>
  <aug>
    <au><snm>Pedregosa</snm><fnm>F</fnm></au>
    <au><snm>Varoquaux</snm><fnm>G</fnm></au>
    <au><snm>Gramfort</snm><fnm>A</fnm></au>
    <au><snm>Michel</snm><fnm>V</fnm></au>
    <au><snm>Thirion</snm><fnm>B</fnm></au>
    <au><snm>Grisel</snm><fnm>O</fnm></au>
    <au><snm>Blondel</snm><fnm>M</fnm></au>
    <au><snm>Prettenhofer</snm><fnm>P</fnm></au>
    <au><snm>Weiss</snm><fnm>R</fnm></au>
    <au><snm>Dubourg</snm><fnm>V</fnm></au>
    <au><cnm>others</cnm></au>
  </aug>
  <source>Journal of Machine Learning Research</source>
  <pubdate>2011</pubdate>
  <volume>12</volume>
  <issue>Oct</issue>
  <fpage>2825</fpage>
  <lpage>-2830</lpage>
</bibl>

<bibl id="B62">
  <title><p>The random subspace method for constructing decision
  forests</p></title>
  <aug>
    <au><snm>Ho</snm><fnm>TK</fnm></au>
  </aug>
  <source>IEEE transactions on pattern analysis and machine
  intelligence</source>
  <publisher>IEEE</publisher>
  <pubdate>1998</pubdate>
  <volume>20</volume>
  <issue>8</issue>
  <fpage>832</fpage>
  <lpage>-844</lpage>
</bibl>

<bibl id="B63">
  <title><p>Support-vector networks</p></title>
  <aug>
    <au><snm>Cortes</snm><fnm>C</fnm></au>
    <au><snm>Vapnik</snm><fnm>V</fnm></au>
  </aug>
  <source>Machine learning</source>
  <publisher>Springer</publisher>
  <pubdate>1995</pubdate>
  <volume>20</volume>
  <issue>3</issue>
  <fpage>273</fpage>
  <lpage>-297</lpage>
</bibl>

<bibl id="B64">
  <title><p>A set of descriptors for identifying the protein--drug interaction
  in cellular networking</p></title>
  <aug>
    <au><snm>Nanni</snm><fnm>L</fnm></au>
    <au><snm>Lumini</snm><fnm>A</fnm></au>
    <au><snm>Brahnam</snm><fnm>S</fnm></au>
  </aug>
  <source>Journal of theoretical biology</source>
  <publisher>Elsevier</publisher>
  <pubdate>2014</pubdate>
  <volume>359</volume>
  <fpage>120</fpage>
  <lpage>-128</lpage>
</bibl>

</refgrp>
} % end of \BMCxmlcomment
% for author-year bibliography (bmc-mathphys or spbasic)
% a) write to bib file (bmc-mathphys only)
% @settings{label, options="nameyear"}

% or include bibliography directly:
% \begin{thebibliography}
% \bibitem{b1}
% \end{thebibliography}

%%%%%%%%%%%%%%%%%%%%%%%%%%%%%%%%%%%
%%                               %%
%% Figures                       %%
%%                               %%
%% NB: this is for captions and  %%
%% Titles. All graphics must be  %%
%% submitted separately and NOT  %%
%% included in the Tex document  %%
%%                               %%
%%%%%%%%%%%%%%%%%%%%%%%%%%%%%%%%%%%

%%
%% Do not use \listoffigures as most will included as separate files

\section*{Figures}
%  \begin{figure}[h!]
 % \caption{\csentence{Sample figure title.}
  %    A short description of the figure content
   %   should go here.}
    %  \end{figure}

%\begin{figure}[h!]
 % \caption{\csentence{Sample figure title.}
  %    Figure legend text.}
   %   \end{figure}

\clearpage
%%%%%%%%%%%%%%%%%%%%%%%%%%%%%%%%%%%

%%                               %%
%% Tables                        %%
%%                               %%
%%%%%%%%%%%%%%%%%%%%%%%%%%%%%%%%%%%

%% Use of \listoftables is discouraged.
%%
\section*{Tables}
\begin{table}[!htb]
\caption{Description of the gold standard datasets \cite{yamanishi2008prediction}\label{tabDataset}}
\begin{center}
\begin{tabular}{l|ccc}
\hline
Dataset & Drugs & Proteins & Positive Interactions\\
\hline
Enzyme&445&664&2926\\
Ion Chanel&210&204&1476\\
GPCR&223&95&635\\
Nuclear Receptor&54&26&90\\
\hline
\end{tabular}
\end{center}
\end{table}

\begin{table}[!htb]

\caption{Summary of evolutionary and structural features used for protein targets and fingerprint features for drugs. The ``Group" column shows different feature groups used in our experiments and will be discussed in a later section. \label{tabFeatSum} }

\begin{center}
\begin{tabular}{llcc}
\hline
\bf Feature Group & \bf Number of Features&Reference&Group\\
\hline
Molecular finger print & 881 & \cite{mousavian2016drug}& A\\
PSSM bigram& 400& \cite{mousavian2016drug}\\
\hline
%Secondary Structure Occurence & 3& This paper\\
Secondary Structure Composition & 3& This paper&B\\
Accessible Surface Area Composition & 1& This paper\\
Torsional Angles Composition & 8& This paper\\
\hline
%Structural Probabilities Composition & 3& This paper\\
Torsional Angles Auto-Covariance& 80& This paper&C\\
Structural Probabilities Auto-Covariance&30& This paper\\
\hline
Torsional Angles bigram& 64& This paper&D\\
Structural Probabilities bigram&9& This paper\\
\hline
Total&1476\\
\hline
\end{tabular}
\end{center}
\end{table}

\begin{table}[!htb]

\caption{Performance of different classifiers on the different datasets in terms of are under Receiver Operating Characteristic (ROC) curve (auROC) and area under precision recall curve (auPR) using different feature group combinations and random under sampling.  \label{tabClassifiers}}
\begin{center}
%\begin{scriptsize}
\begin{tabular}{c|c|c|c|c}
\hline
Dataset & Feature Combination& Classifier &auPR&auROC\\
\hline
\hline
enzymes&A&AdaBoost&0.54&0.9530\\
\cline{3-5}
&&Random Forest&0.43&0.9457\\
\cline{3-5}
&&SVM&\bf 0.64&\bf 0.9647\\
\cline{2-5}
&A,B&AdaBoost&\bf 0.51&0.9431\\
\cline{3-5}
&&Random Forest&0.49&0.9445\\
\cline{3-5}
&&SVM&0.48&\bf 0.9502\\
\cline{2-5}
&A,B,C&AdaBoost&\bf 0.66&\bf 0.9638\\
\cline{3-5}
&&Random Forest&0.48&0.9334\\
\cline{3-5}
&&SVM&0.41&0.9360\\
\cline{2-5}
&A,B,C,D&AdaBoost&\bf 0.65&\bf 0.9689\\
\cline{3-5}
&&Random Forest&0.50&0.9493\\
\cline{3-5}
&&SVM&0.63&0.9628\\
\cline{2-5}
\hline
\hline
ion channels&A&AdaBoost&\bf 0.36&0.9271\\
\cline{3-5}
&&Random Forest&0.33&0.9232\\
\cline{3-5}
&&SVM&0.25&\bf 0.9467\\
\cline{2-5}
&A,B&AdaBoost&\bf 0.33&0.9191\\
\cline{3-5}
&&Random Forest&0.30&0.8898\\
\cline{3-5}
&&SVM&0.23&\bf 0.9213\\
\cline{2-5}
&A,B,C&AdaBoost&\bf 0.34&0.9202\\
\cline{3-5}
&&Random Forest&0.31&0.8734\\
\cline{3-5}
&&SVM&0.23&\bf 0.9213\\
\cline{2-5}
&A,B,C,D&AdaBoost&\bf 0.43&\bf 0.9369\\
\cline{3-5}
&&Random Forest&0.40&0.9234\\
\cline{3-5}
&&SVM&0.14&0.6723\\
\cline{2-5}
\hline
\hline
GPCRs&A&AdaBoost&\bf 0.29&\bf 0.8856\\
\cline{3-5}
&&Random Forest&0.23&0.8743\\
\cline{3-5}
&&SVM&0.18&0.7832\\
\cline{2-5}
&A,B&AdaBoost&\bf 0.29&\bf 0.8834\\
\cline{3-5}
&&Random Forest&0.22&0.8698\\
\cline{3-5}
&&SVM&0.15&0.7802\\
\cline{2-5}
&A,B,C&AdaBoost&\bf 0.35&\bf 0.9116\\
\cline{3-5}
&&Random Forest&0.31&0.9034\\
\cline{3-5}
&&SVM&0.15&0.7945\\
\cline{2-5}
&A,B,C,D&AdaBoost&\bf 0.31&0.9128\\
\cline{3-5}
&&Random Forest&0.30&\bf 0.9168\\
\cline{3-5}
&&SVM&0.21&0.7896\\
\cline{2-5}
\hline
\hline
nuclear receptors&A&AdaBoost&\bf 0.41&\bf 0.8145\\
\cline{3-5}
&&Random Forest&0.23&0.7519\\
\cline{3-5}
&&SVM&0.19&0.7898\\
\cline{2-5}
&A,B&AdaBoost&\bf 0.43&\bf 0.7969\\
\cline{3-5}
&&Random Forest&0.29&0.7723\\
\cline{3-5}
&&SVM&0.20&0.6789\\
\cline{2-5}
&A,B,C&AdaBoost&\bf 0.36&\bf 0.7590\\
\cline{3-5}
&&Random Forest&0.21&0.7234\\
\cline{3-5}
&&SVM&0.21&0.6971\\
\cline{2-5}
&A,B,C,D&AdaBoost&\bf 0.33&\bf 0.7946\\
\cline{3-5}
&&Random Forest&0.29&0.7145\\
\cline{3-5}
&&SVM&0.20&0.7287\\
\cline{2-5}
\hline
\end{tabular}
%\end{scriptsize}
\end{center}
\end{table}

\begin{table}[!htb]
\caption{Parameters of AdaBoost Algorithm used with decision tree as weak classifier along with different balancing methods on four datasets \label{tabParams}}
\begin{center}
%\begin{scriptsize}
\begin{tabular}{c|c | c c c c }

	\hline
balancing& & Max  & Min sample   & Min samples &  \\
method&Dataset& depth&split&Leaf &Criterion 	\\
	\hline
	\hline
random&	enzymes& 100 & 16 & 1 & Gini impurity \\
	\cline{2-6}
&	ion channels& 8  & 4 & 1 & Gini impurity\\
	\cline{2-6}
&	GPCRs& 6 & 3 & 1 & Gini impurity \\
	\cline{2-6}
&	nuclear receptors& 5 & 7 & 2 & Gini impurity \\
	\hline
	\hline
	clustered&enzymes& 110 & 1 & 1 & Gini impurity \\
	\cline{2-6}
	&ion channels& 9  & 2 & 1 & Gini impurity\\
	\cline{2-6}
	&GPCRs& 6 & 3 & 1 & Gini impurity \\
	\cline{2-6}
	&nuclear receptors& 150 & 2 & 1 & Gini impurity \\
	\hline
\end{tabular}
%\end{scriptsize}
\end{center}

\end{table}

\begin{table}[!htb]

\caption{Performance of Adaboost classifier on different datasets in terms of are under Receiver Operating Characteristic (ROC) curve (auROC) and area under precision recall curve (auPR) using different feature group combinations and balancing methods.  \label{tabFeatBalance}}
\begin{center}
%\begin{small}
\begin{tabular}{c|c|c|c|c}
\hline
Dataset & Feature Combination& Balancing Method &auPR&auROC\\
\hline
\hline
enzymes&A&random&0.54&0.9530\\
\cline{3-5}
&&clustered&0.58&0.9493\\
\cline{2-5}
&A,B&random&0.51&0.9431\\
\cline{3-5}
&&clustered&0.59&0.9353\\
\cline{2-5}
&A,B,C&random&0.66&0.9638\\
\cline{3-5}
&&clustered&0.63&0.9577\\
\cline{2-5}
&A,B,C,D&random&0.65&\bf 0.9689\\
\cline{3-5}
&&clustered&\bf 0.68&0.9598\\
\hline
\hline
ion channels&A&random&0.36&0.9271\\
\cline{3-5}
&&clustered&0.38&0.8982\\
\cline{2-5}
&A,B&random&0.33&0.9191\\
\cline{3-5}
&&clustered&0.41&0.8902\\
\cline{2-5}
&A,B,C&random&0.34&0.9202\\
\cline{3-5}
&&clustered&0.45&0.9021\\
\cline{2-5}
&A,B,C,D&random&0.43&\bf 0.9369\\
\cline{3-5}
&&clustered&\bf 0.48&0.9051\\
\hline
\hline
GPCRs&A&random&0.29&0.8856\\
\cline{3-5}
&&clustered&0.48&0.9189\\
\cline{2-5}
&A,B&random&0.29&0.8834\\
\cline{3-5}
&&clustered&0.49&0.8968\\
\cline{2-5}
&A,B,C&random&0.35&0.9116\\
\cline{3-5}
&&clustered&\bf 0.50&0.8890\\
\cline{2-5}
&A,B,C,D&random&0.31&0.9128\\
\cline{3-5}
&&clustered&0.48&\bf 0.9322\\
\hline
\hline
nuclear receptors&A&random&0.41&0.8145\\
\cline{3-5}
&&clustered&0.79&0.9270\\
\cline{2-5}
&A,B&random&0.43&0.7969\\
\cline{3-5}
&&clustered&0.32&0.8715\\
\cline{2-5}
&A,B,C&random&0.36&0.7590\\
\cline{3-5}
&&clustered&0.57&0.8935\\
\cline{2-5}
&A,B,C,D&random&0.33&0.7946\\
\cline{3-5}
&&clustered&\bf 0.79&\bf 0.9285\\
\hline
\end{tabular}
%\end{small}
\end{center}
\end{table}

\begin{table}[!htb]
\caption{Comparison of the performance of {\methodname} on the four bechmark gold datasets in terms on area under receiver operating characteristic curve (auROC) with other state-of-the-art methods.\label{tabMain}}
\begin{center}
%\begin{footnotesize}
\begin{tabular}{l|c|c|c|c|c|c|c|c}
\hline
Dataset & DBSI & KBMF2K & NetCBP & Yamanishi & Yamanishi & Wang  & Mousavian & {\methodname}\\
&\cite{cheng2012prediction}&\cite{gonen2012predicting}&\cite{chen2013semi}&et al. \cite{yamanishi2008prediction}&et al. \cite{yamanishi2010drug}&et al. \cite{wang2013drug}&et al. \cite{mousavian2016drug}&\\
\hline
enzymes&0.8075&0.8320&0.8251&0.904&0.8920&0.8860&0.9480&\bf 0.9689\\
ion channels&0.8029&0.7990&0.8034&0.8510&0.8120&0.8930&0.8890&\bf 0.9369\\
GPCRs&0.8022&0.8570&0.8235&0.8990&0.8270&0.8730&0.8720&\bf 0.9322\\
nuclear receptors&0.7578&0.8240&0.8394&0.8430&0.8350&0.8240&0.8690&\bf 0.9285\\
\hline
\end{tabular}
%\end{footnotesize}
\end{center}

\end{table}

\begin{table}[!htb]
\caption{Comparison of the performance of {\methodname} on the four benchmark gold datasets in terms on area under precision-recall curve (auPR) with the state-of-the-art method in \cite{mousavian2016drug}.\label{tabMainPR}}
\begin{center}
%\begin{footnotesize}
\begin{tabular}{l|c|c|c|c}
\hline
Predictor & enzymes & ion channels & GPCRs & nuclear receptors \\
\hline
Mousavian et al. \cite{mousavian2016drug} &0.546&0.390&0.282&0.411\\
{\methodname} &\bf 0.680&\bf 0.480&\bf 0.500&\bf 0.790\\
\hline
\end{tabular}
%\end{footnotesize}
\end{center}
\end{table}

\begin{table}
\caption{Specificity, Sensitivity, Precision, MCC and F1 score for four datasets as achieved by {\methodname} using different feature groups. \label{tabOthers}}
\begin{tabular}{  l | l | l | l | l | l | l  }
	\hline
	dataset&Feature Group & Specificity & Sensitivity  & Precision  & MCC & F1 score   \\ \hline
	enzymes 	& A 	  & 0.83& 0.9 	& 0.05 	& 0.1962& 0.10 	\\ 
	\cline{2-7}
	& A,B 	  & 0.82& 0.89 	& 0.05 	& 0.1812& 0.09 	\\ \cline{2-7}
	& A,B,C   & 0.83& 0.87 	& 0.05 	& 0.1762& 0.09 	\\ \cline{2-7}
	& A,B,C,D & 0.85&0.85 	& 0.15 	& 0.1889& 0.10 	\\ \hline
	ion channels 		& A  	  & 0.78& 0.81 	& 0.13 	& 0.2615& 0.22 	\\ \cline{2-7}
	& A,B 	  & 0.78& 0.84 	& 0.14 	& 0.256 & 0.24 	\\ \cline{2-7}
	& A,B,C   & 0.8 & 0.86 	& 0.12 	& 0.2980& 0.20 	\\ \cline{2-7}
& A,B,C,D & 0.78& 0.84 	& 0.13 	& 0.2913& 0.20 	\\ \hline
	GPCRs 	& A 	  & 0.78& 0.84 	& 0.12 	& 0.254 & 0.20 	\\ \cline{2-7}
	& A,B 	  & 0.8 & 0.85 	& 0.11 	& 0.2760& 0.20 	\\ \cline{2-7}
	& A,B,C   & 0.79& 0.89 	& 0.11 	& 0.2797& 0.19 	\\ \cline{2-7}
	& A,B,C,D & 0.8 & 0.84 	& 0.11 	& 0.2647& 0.19 	\\ \hline
	nuclear receptors & A 	  & 0.85& 0.91 	& 0.16	& 0.2141& 0.27 	\\ \cline{2-7}
	& A,B 	  & 0.77& 0.88 	& 0.11 	& 0.2154& 0.19 	\\ \cline{2-7}
	& A,B,C   & 0.81& 0.88 	& 0.12 	& 0.1798& 0.20 	\\ \cline{2-7}
	& A,B,C,D & 0.92& 0.87 	& 0.14 	& 0.2253& 0.24 	\\ \hline
\end{tabular}
\end{table}

%%%%%%%%%%%%%%%%%%%%%%%%%%%%%%%%%%%
%%                               %%
%% Additional Files              %%
%%                               %%
%%%%%%%%%%%%%%%%%%%%%%%%%%%%%%%%%%%
\clearpage
\section*{Additional Files}
  \subsection*{Additional file 1 --- AdditionalFile1.xlsx}
    AdditionalFile1.xlsx contains the new drug-target predictions from each of the datasets. For each dataset, 10 top predictions from the false negatives with highest probability scores are reported.

%\end{backmatter}
\end{document}